\documentclass[sigconf]{acmart}

\usepackage[utf8]{inputenc} 
\usepackage[T1]{fontenc}    
\usepackage{hyperref}       
\usepackage{url}            
\usepackage{amsfonts}       
\usepackage{nicefrac}       
\usepackage{microtype}      
\usepackage{xcolor}         
\usepackage{graphicx,subcaption}
\usepackage{amsmath}
\usepackage{enumitem}
\usepackage{multirow}
\usepackage{balance}
\usepackage{booktabs}

\copyrightyear{2026}
\acmYear{2026}
\setcopyright{cc}
\setcctype{by}
\acmConference[CHI '26]{Proceedings of the 2026 CHI Conference on Human Factors in Computing Systems}{April 13--17, 2026}{Barcelona, Spain}
\acmBooktitle{Proceedings of the 2026 CHI Conference on Human Factors in Computing Systems (CHI '26), April 13--17, 2026, Barcelona, Spain}
\acmPrice{}
\acmDOI{10.1145/3772318.3791322}
\acmISBN{979-8-4007-2278-3/2026/04}

\newcommand{\xref}[1]{\S\ref{#1}}

\newcommand{\squishlist}{\begin{itemize}[itemsep=1pt,parsep=2pt,topsep=3pt,partopsep=0pt,leftmargin=0em, itemindent=1em,labelwidth=1em,labelsep=0.5em]}
\newcommand{\squishend}{\end{itemize}}

\begin{document}

\title{{VueBuds: Visual Intelligence with  Wireless Earbuds}}

 \author{Maruchi Kim}
 \affiliation{Paul G. Allen School,  \\University of Washington, WA  
 \country{USA}
 }
 \email{mkimhj@cs.washington.edu }

 \author{Rasya Fawwaz}
 \affiliation{Electrical \& Computer Engineering, University of Washington,  WA
  \country{USA}
 }
 \email{fawwa001@uw.edu}

  \author{Zhi Yang Lim}
 \affiliation{Paul G. Allen School, \\University of Washington,  WA
  \country{USA}
 }
 \email{zylim@cs.washington.edu}

  \author{Brinda Moudgalya}
 \affiliation{Electrical \& Computer Engineering, University of Washington, WA
  \country{USA}
 }
 \email{brindam@uw.edu}

  \author{Hexi Wang}
 \affiliation{Electrical \& Computer Engineering, University of Washington, WA
  \country{USA}
 }
 \email{whx0627@uw.edu}

  \author{Yuanhao Zeng}
 \affiliation{Electrical \& Computer Engineering, University of Washington, WA
  \country{USA}
 }
 \email{yhzeng9@uw.edu}
 
 \author{Shyamnath Gollakota}
 \affiliation{Paul G. Allen School,\\ University of Washington,  WA
   \country{USA}
 }
\email{gshyam@cs.washington.edu}

\renewcommand{\shortauthors}{Kim, Fawwaz, Lim, Moudgalya, Wang, Zeng and Gollakota}



\begin{teaserfigure}
  \includegraphics[width=\textwidth]{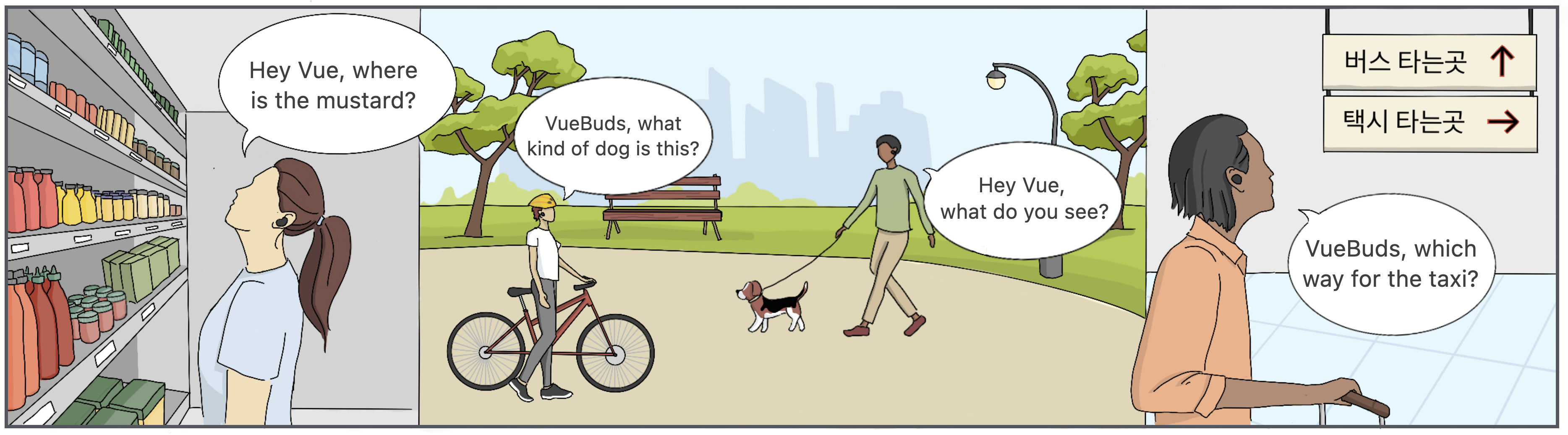}
  \vskip -0.1in
  \caption{{\bf Applications of VueBuds.} \textmd {Our camera-integrated wireless earbuds enable natural language queries for everyday visual tasks such as locating items in a store, identifying objects, obtaining scene-level descriptions, and interpreting foreign text for navigation.}}
  \Description{The illustration shows four everyday scenarios where people use camera-equipped earbuds called VueBuds to ask visual questions. In a grocery store, a shopper looks at shelves of condiments and asks, “Hey Vue, where is the mustard?” In a park, a person on a bicycle looks at a nearby dog and asks, “VueBuds, what kind of dog is this?” Another park scene shows a person walking a dog while asking, “Hey Vue, what do you see?” At a transit area with Korean signs pointing to bus and taxi stands, a traveler asks, “VueBuds, which way for the taxi?” The figure demonstrates potential use cases for wearable visual AI in shopping, outdoor exploration, accessibility, and travel.}
  \label{fig:teaser}
\end{teaserfigure}

\begin{abstract}

Despite their ubiquity, wireless earbuds remain audio-centric due to size and power constraints. We present VueBuds, the first camera-integrated wireless earbuds for egocentric vision, capable of operating within stringent power and form-factor limits. Each VueBud embeds a camera into a Sony WF-1000XM3 to stream visual data over Bluetooth to a host device for on-device vision language model (VLM) processing. We show analytically and empirically that while each camera's field of view is partially occluded by the face, the combined binocular perspective provides comprehensive forward coverage. By integrating VueBuds with VLMs, we build an end-to-end system for real-time scene understanding, translation, visual reasoning, and text reading; all from low-resolution monochrome cameras drawing under 5mW through on-demand activation. Through online and in-person user studies with 90 participants, we compare VueBuds against smart glasses across 17 visual question-answering tasks, and show that our system achieves response quality on par with Ray-Ban Meta. Our work establishes low-power camera-equipped earbuds as a compelling platform for visual intelligence, bringing rapidly advancing VLM capabilities to one of the most ubiquitous wearable form factors. Video demos at: {\textcolor{blue}{\url{https://vuebuds.cs.washington.edu/}}}.

\end{abstract}



\maketitle

\section{Introduction}

The emergence of large language models has  transformed human-computer interaction, enabling natural language conversations with intelligent systems across diverse applications~\cite{bubeck2023sparks,nature11}. This has accelerated further with the integration of visual intelligence, giving rise to vision language models (VLMs) that can comprehend and reason about images alongside text~\cite{llava,flamingo}. These advances have found rapid adoption in consumer devices, with smartphones leveraging on-device visual intelligence for enhanced photography and accessibility ~\cite{BixbyVision}, while smart glasses like Ray-Ban Meta demonstrate the potential of wearable visual computing in everyday life ~\cite{rayban}.

Despite advances in integrating visual intelligence into mobile and wearable devices, wireless earbuds remain largely limited to audio-centric functionality. Current designs incorporate low-power peripherals such as microphones, inertial measurement units, biometric sensors, and speakers~\cite{imuposer,earbit}. The absence of visual capabilities in this ubiquitous form factor represents a notable gap in the wearable ecosystem, particularly since wireless earbuds have orders-of-magnitude greater commercial adoption, with an estimated user base 150–200x larger than that of smart glasses~\cite{glassmarket,earbudmarket}.

We introduce VueBuds, the first wireless earbud system integrating low-power cameras with vision language model interaction. VueBuds allow users to capture visual context from their environment and engage with vision language models through a familiar, everyday wearable platform, without requiring specialized eyewear. Our binaural system integrates dual forward-facing cameras, leveraging binocular vision to overcome facial occlusions and capture the wearer's egocentric view.

{Achieving this requires addressing three core research questions:}
\squishlist

\item {{\it RQ1: Can truly wireless earbuds support camera hardware within strict size, weight, and power (SWaP) constraints?} Camera sensors and visual processing consume far more power than typical earbud components, raising questions about whether earbuds can support cameras while preserving acceptable battery life and form factor. Furthermore, camera data demands significantly higher bandwidth than audio, making it unclear whether the wireless protocols used in earbuds can effectively transmit a binocular visual stream. }

\item { {\it RQ2: To what extent can cameras positioned at ear level provide robust egocentric views for visual perception and interaction?} Unlike smart glasses, where cameras are unobstructed and align closely with the user's eyes, ear-level cameras have a posterior-lateral offset. This placement introduces potential facial occlusion, raising an unresolved question of whether such a vantage point can support effective egocentric vision.}



\item  {{\it RQ3: Can a fully wireless, Bluetooth-based pipeline support real-time multimodal interaction with vision–language models?} Answering user queries (e.g., ``Where are my keys?'') requires capturing contextually relevant imagery, streaming it via low-bandwidth Bluetooth,\footnote{Bluetooth consumes far less power than Wi-Fi but offers significantly lower bandwidth. Smart glasses can support Wi-Fi due to their larger batteries and form factor.} performing multimodal inference using an on-device vision–language model, and synthesizing an audio response. Meeting real-time latency constraints across this end-to-end pipeline remains an open systems challenge.}


\begin{figure}[t!]
\vskip -0.1in
\centering
\includegraphics[width=\linewidth]{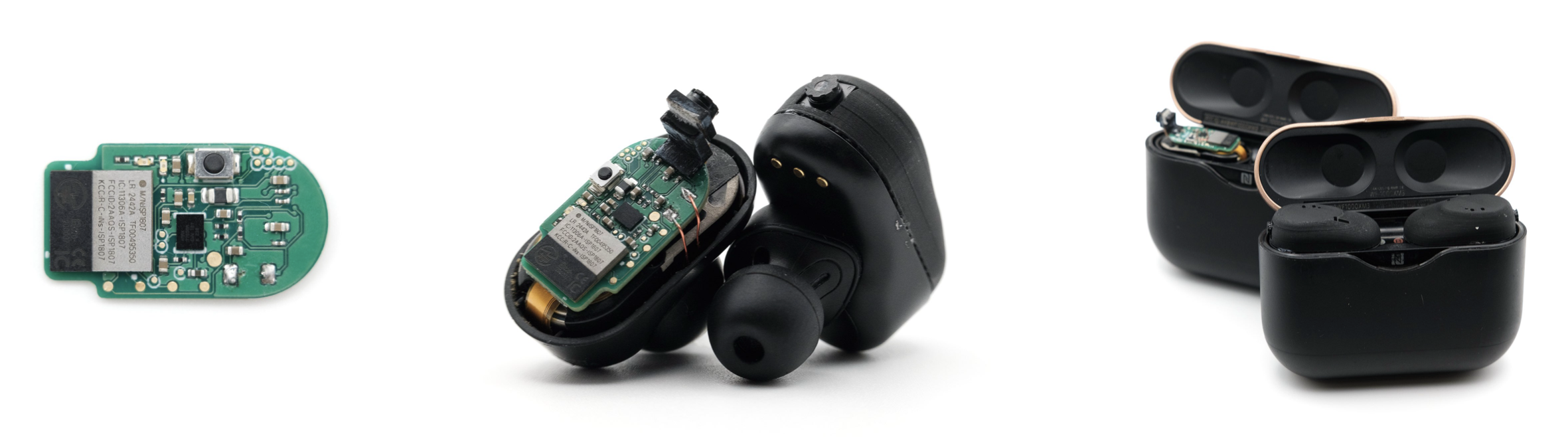}
\vskip -0.1in

\caption{{\bf VueBuds integrated with Sony wireless earbuds. \textmd{The custom camera module (left) is powered directly from the earbud battery, with 3D-printed enclosures (middle) enabling forward-facing capture. VueBuds charge via the original case (right).}}}

\Description{This image contains the VueBuds hardware integrated with Sony WF-1000XM3 wireless earbuds. The left image shows the printed circuit board of the custom camera module with the BLE microcontroller, PMIC, and camera sensor. The middle image shows the camera module attached to the earbud using a 3D-printed enclosure for forward-facing capture. The right image shows the earbuds placed inside the original Sony charging case, demonstrating compatibility with the charging system.}
\label{fig:vuebuds}
\vskip -0.15in
\end{figure}


\squishend

{We address these questions through four key contributions:}
\squishlist

\item {{\bf Camera-integrated wireless earbud hardware.} We develop the first dual-camera earbud prototype, maintaining practical size, weight, and power expected in this form factor (Fig.~\ref{fig:vuebuds}). The custom camera module attaches to commodity Sony WF-1000XM3 earbuds and operates at under 5~mW (\xref{sec:poweranalysis}), adding only 11–14\% battery overhead even under extreme usage of 60 visual queries per hour.}

 
\item {{\bf Binocular vision for facial occlusion.} We introduce an ear-level egocentric capture system that leverages dual viewpoints to resolve unilateral obstructions. Analytical modeling and empirical validation demonstrate that this approach significantly reduces blind spots, maintaining occlusion depths well below the Harmon distance threshold, the practical limit for supporting effective egocentric interaction (\xref{sec:modeling}).}

\begin{table*}[t!]
\centering
\caption{{\bf {Comparison of egocentric visual wearable systems.}} {\textmd{{XR headsets (e.g., Vision Pro) offer eye-level alignment with rich deictic cues (gaze + hand tracking) but are bulky and isolating. Smart glasses (e.g., Ray-Ban Meta) resemble ordinary eyewear, but lack gaze tracking. Body-worn cameras (e.g. AI Pin, GoPro) provide torso-level viewpoints with optional projection feedback. VueBuds utilize ear-level cameras in a compact, familiar form factor. Beyond these technical trade-offs, device adoption is shaped by social factors: earbuds represent the most widely adopted wearable category, 
with over 340 million units shipped annually compared to approximately 2 million smart glasses~\cite{idc2024wearables}, suggesting that ear-worn devices may face lower barriers for everyday use.}}}}

\label{tab:system-comparison}
\begin{tabular}{lccccc}
\hline
\addlinespace[1.5pt]
\textbf{\shortstack{Form Factor}} & \textbf{\shortstack{Viewpoint Alignment}} & \textbf{\shortstack{Deictic Cues}} & \textbf{\shortstack{Privacy Signaling}} & \textbf{\shortstack{Feedback Modality}} \\
\addlinespace[1.5pt]
\hline

XR Headsets & Eye-level & Gaze + Hand + Voice & High & Visual + Audio \\
Smart Glasses & Eye-Level & Hand + Voice & Low-Medium & Visual + Audio \\
Body-worn Cameras & Torso-level & Hand + Voice & Medium-High & Projector + Audio \\
\textbf{VueBuds (Ours)} & Ear-Level & Hand + Voice & Low & Audio \\
\hline
\end{tabular}
\end{table*}

\item {{\bf End-to-end system optimizations for real-time operation and VLM integration.} In~\xref{sec:vuebudhardware}, we detail optimizations that maximize wireless throughput and minimize latency for concurrent binaural video streaming. We demonstrate that VLMs such as Qwen2.5-VL can support robust scene understanding, translation, and text reading despite the earbuds' low-resolution, monochrome imagery. comparing input strategies, we find that stitching L/R images outperforms separate processing by eliminating visual redundancy, improving time-to-first-token (TTFT) by 46\%. VueBuds achieves an end-to-end latency of under 3 seconds, with identified optimizations (\xref{sec:latencyresults}) capable of reducing this to 1.14s.}

\item {{\bf User studies and comparative evaluation with smart glasses.}  Across two user studies ($n=90$), we assess real-world feasibility and user acceptance (\xref{sec:studies}). An online study ($n=74$) shows that earbuds are far more widely adopted than regular glasses (93.3\% vs. 62.7\%) and that VueBuds+Qwen2.5-VL delivers visual question answering (VQA) performance across 17 tasks comparable to Ray-Ban Meta (MOS: 3.33 vs. 3.32). An in-person study ($n=16$) demonstrates strong real-time performance in object recognition (82.5\%), optical character recognition (OCR) (94.3\%), and translation (83.8\%). Participants also reported comfort similar to everyday earbuds and perceived broad applicability.}

\squishend

\vskip 0.05in\noindent{{\bf Key findings.} Our results demonstrate that:  (1) integrating cameras into earbuds is feasible within strict SWaP constraints; (2) binocular ear-level capture effectively mitigates facial occlusion to provide egocentric views; (3) our fully-wireless pipeline with modern VLMs can operate effectively on low-resolution, monochrome images in real time. Further, VueBuds is  perceived as highly accessible for multimodal interaction, achieving utility competitive with Ray-Ban Meta glasses. Together, these findings establish earbuds as a promising platform for egocentric visual intelligence applications.}

\section{Related work}

{We position VueBuds within three interconnected research areas. First, we examine existing wearable visual intelligence systems, from smart glasses to AR headsets, and their approaches to situated interaction and context-aware assistance. Second, we review the evolution of ear-worn sensing platforms, identifying the absence of visual capabilities as a key gap. Finally, we address the cross-cutting challenges of power consumption and privacy that any wearable camera system must navigate.}

\vskip 0.05in\noindent{\it Head-mounted visual wearables.} Existing visual wearable systems have primarily focused on glasses and head-mounted cameras~\cite{chiglasses,10.1145/3654777.3676379}. For example, Pupil Invisible glasses from Pupil Labs provide calibration-free gaze tracking~\cite{pupil}. {CAPturAR ~\cite{CapturAR} uses a customized AR head-mounted device to author context-aware applications by referencing users' previous activities. GazePointAR~\cite{gazepointar} combines eye gaze, pointing gestures, and conversation history to disambiguate speech queries in augmented reality.} In the consumer market, products such as Ray-Ban Meta~\cite{rayban} and Google Glass~\cite{googleglasses} feature forward-facing cameras for photo/video capture and AI-based visual question answering (VQA), {though Google Glass faced negative press due to privacy concerns} \cite {googleglassprivacy}. Other commercial offerings like the XREAL Aura ~\cite{xreal} focus on display functionality rather than visual intelligence, serving as a wired secondary monitors for productivity. {In parallel with these commercial systems, the advent of large language models has sparked recent research in exploring multimodal interaction techniques that combine visual perception with user intent. GesPrompt ~\cite{GesPrompt} demonstrates how co-speech gestures enable more precise object selection in vision language model interactions, allowing users to specify a specific object while pointing. In contrast to these prior works that focus on face-mounted visual wearables, ear-level cameras for general visual intelligence remain commercially unavailable and underexplored in the research community, despite the widespread adoption of wireless earbuds as a wearable platform.  }



\vskip 0.05in\noindent{\it  Ear-worn sensor platforms and hearables.} Modern hearables have evolved to incorporate a variety of sensors, including microphones, inertial measurement units (IMUs), and photoplethysmography (PPG) sensors, enabling health monitoring and motion tracking~\cite{hu2025surveyearabletechnologytrends,earablegoogle,act1,imuposer,eargate}. Prior work has explored diverse activity recognition and health-tracking applications using these sensor platforms~\cite{chi22-ultrasonic,infection,oae,mobisys21,teeth}. Recently, several earable platforms have been developed that integrate multiple of these sensors~\cite{esense,10.1145/3544793.3563415,10.1145/3712069,chatterjee2022clearbuds,neuralaids}, but none have incorporated cameras. These works demonstrate the feasibility of compact, ear-worn devices for rich physiological and facial movement sensing, yet they generally lack egocentric vision capabilities. The most closely related work~\cite{earauthcam} uses an inward-facing camera to capture ear images for authentication, rather than for general visual intelligence, and relies on wired hardware with an evaluation board that does not meet practical power or size constraints. Similarly,~\cite{c-face} employs cameras to reconstruct facial expressions, not for LLM integration, and, like~\cite{earauthcam}, is not wireless.  Though rumored to be in development in tech media~\cite{rumor}, no camera-integrated wireless earbuds have been publicly released.

\vskip 0.05in\noindent{\it  Low-power camera systems. } Prior work has explored several approaches to reducing the power consumption of vision systems, including the design of more power-efficient camera sensors~\cite{camerapower1,camerapower2,michigan}, mixed-signal vision integrated circuits~\cite{mixed2}, and specialized processors~\cite{processors1,processors2}. Low-power wireless video has been demonstrated in various contexts, for example, Bluetooth-based video streaming for robots and sensor systems~\cite{sciencerobotics_uw,neuricam}, as well as backscatter techniques and low-power machine learning algorithms~\cite{sachin1,batteryvideo1,batteryvideo2,lee2025hypercamlowpoweronboardcomputer}. IRIS~\cite{iris} further extends this space with a vision-enabled ring integrated with object detection. In contrast to these efforts, our work  designs and demonstrates the first vision-enabled wireless earbud system for real-time interaction with VLMs.

\section{System Design}
We design VueBuds to integrate visual intelligence into wireless earbuds while preserving the familiar form factor that makes them widely adopted. This section describes our hardware platform, binocular vision approach for capturing the user's forward-facing perspective, and end-to-end system integration with on-device vision language models.

\begin{figure}[t!]
\centering
    \includegraphics[width=0.48\textwidth]{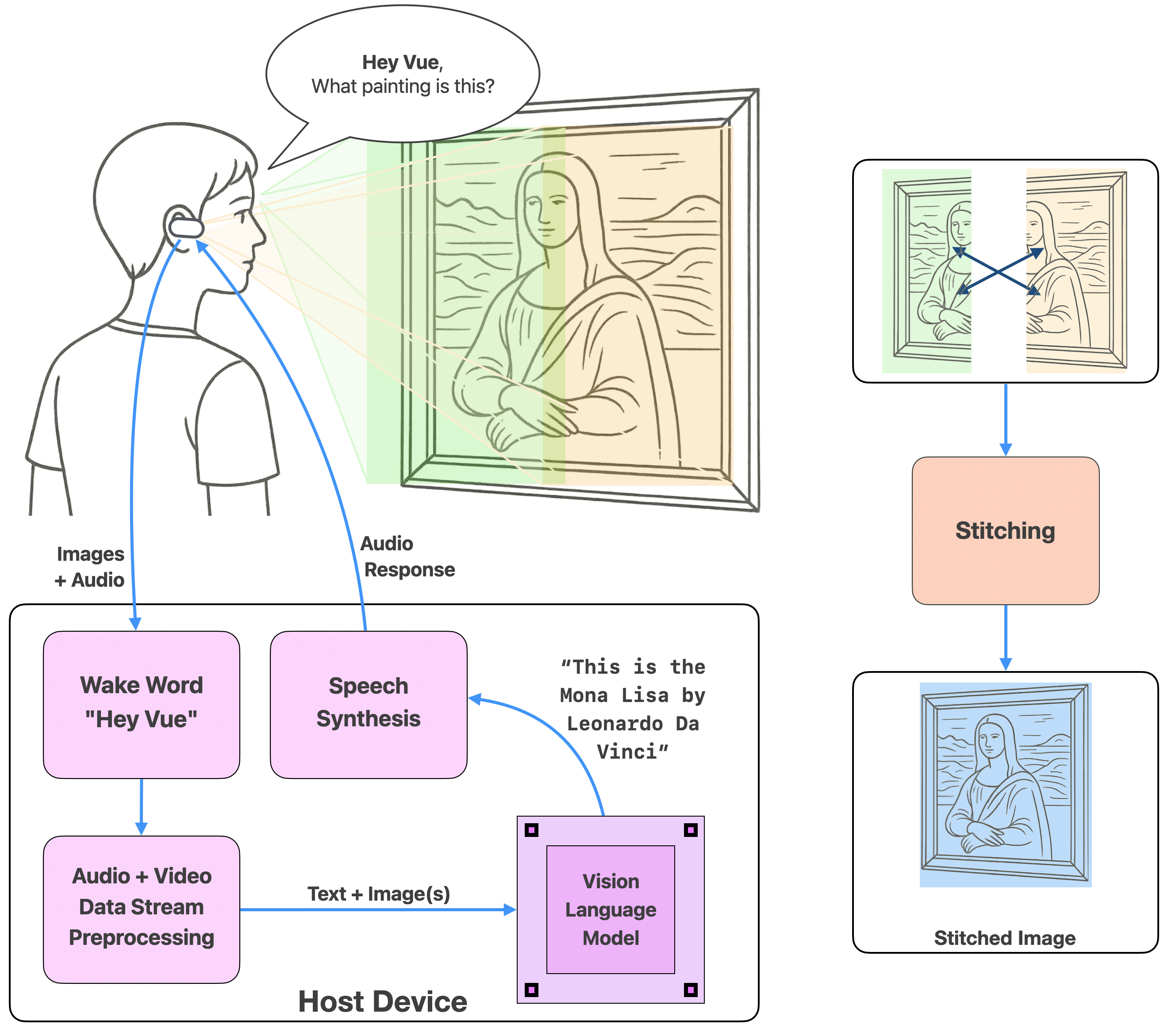}
    \vskip -0.15in
\caption{{\bf VueBuds system overview. \textmd{VueBuds utilizes a vision language model to process the user's voice and binocular camera data for real-time multimodal interaction.}}}
\Description{The diagram illustrates how camera-equipped earbuds (VueBuds) answer a user’s visual question using a vision-language model. On the left, a person wearing VueBuds looks at the Mona Lisa painting and asks aloud, “Hey Vue, what painting is this?” The earbuds capture both images and audio. At the bottom, a flow diagram labeled Host Device shows the processing pipeline: 1)  Wake-Word “Hey Vue” triggers the system. 2)  Audio + Video Data Stream Preprocessing prepares the input. 3) The processed text + image(s) are sent to a Vision Language Model, which generates an answer: “This is the Mona Lisa by Leonardo Da Vinci.” 4)  Speech Synthesis converts the text answer into an audio response, which is delivered back to the user through the earbuds. On the right, a parallel diagram shows the visual processing step: stereo images from the earbuds are combined through Stitching into a single stitched image that represents the full field of view. Together, the figure demonstrates how multimodal inputs (audio and video) flow through the system to provide real-time spoken responses to user queries. }
\label{fig:system-overview}
\vskip -0.35in
\end{figure}

\subsection{VueBuds Hardware} \label{sec:vuebudhardware}

\subsubsection{Hardware Overview}
Our custom hardware design integrates an ultra-low-power Himax HM01B0 CMOS image sensor (1/11" format, 324x324 pixel array), an Analog Devices MAX77650 power management integrated circuit, and a Nordic nRF52840 Bluetooth Low Energy system-on-chip. The design includes status LEDs and a single-pole single-throw switch for debugging and development purposes.

\begin{figure}[t!]
    \centering
    \includegraphics[width=\linewidth]{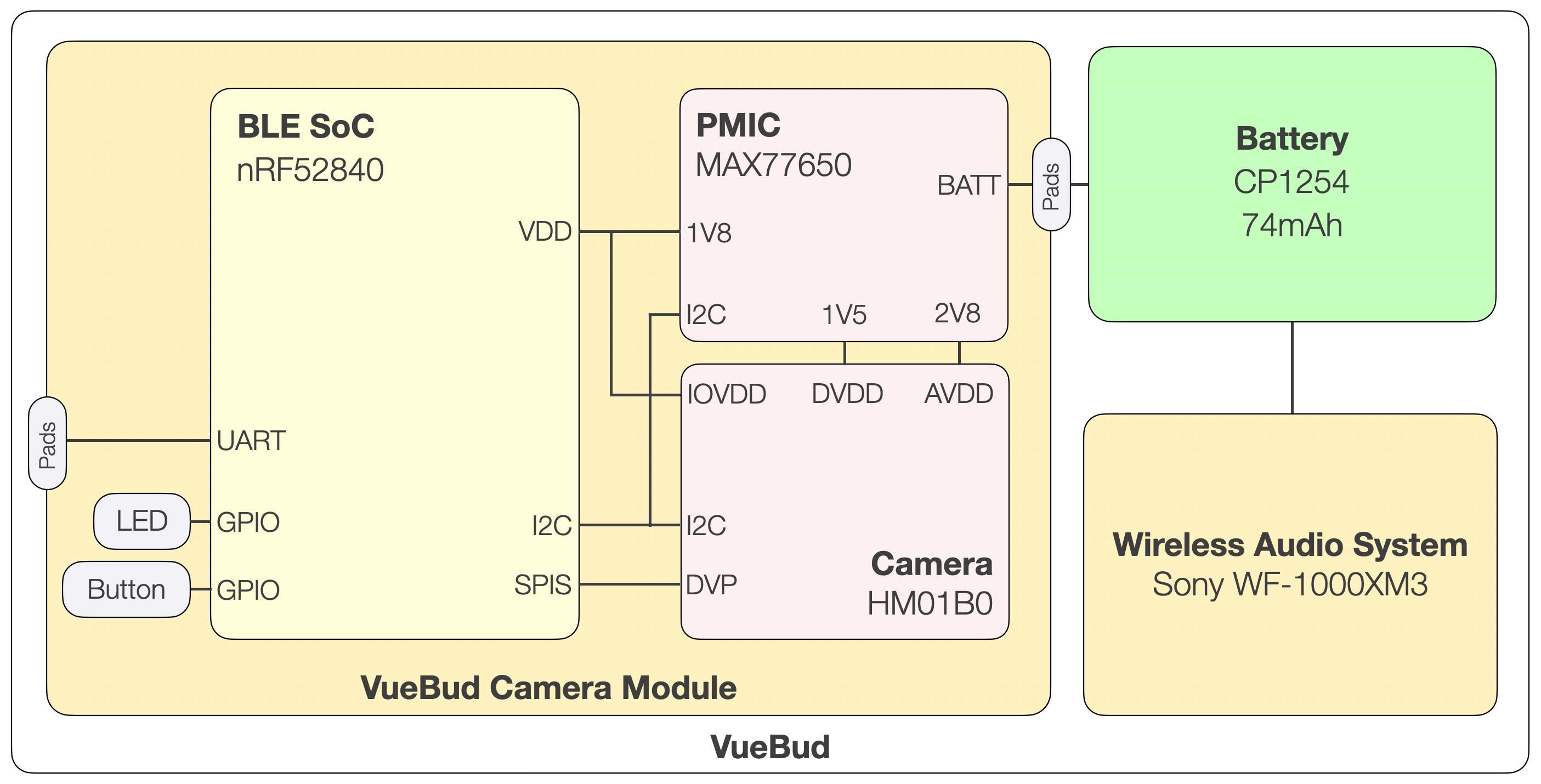}
    \caption{{\bf Hardware block diagram.} \textmd { Each VueBud integrates a Himax CMOS Imaging Camera and a Power Management Integrated Circuit (PMIC). Power is delivered through the onboard battery inside a Sony WF-1000XM3 wireless earbud. }} 
    \Description{The diagram shows the hardware architecture of a VueBud device, highlighting its main electronic components and connections. On the left is the VueBud Camera Module, which includes: 1) A BLE SoC (nRF52840) that provides wireless communication and interfaces for UART, GPIO, I2C, and SPIS. It connects to debug pads, an LED, and a button. 2)  A Power Management Integrated Circuit (PMIC, MAX77650) that regulates multiple voltages (1V8, 1V5, 2V8) for the system. 3) A Camera (Himax HM01B0 CMOS imaging sensor) connected to the PMIC. On the right side: A Battery (CP1254, 74 mAh) supplies power to the system. A Wireless Audio System (Sony WF-1000XM3) integrates audio functionality with the camera module. Connections between blocks show how power, control signals, and data flow through the device. The caption explains that each VueBud integrates a Himax CMOS imaging camera and PMIC, powered by the onboard battery of a Sony WF-1000XM3 wireless earbud. Buttons, LEDs, and UART pads are provided for debugging.}
   \vskip -0.25in
    \label{fig:blockdiagram}
\end{figure}

The custom companion PCB is integrated into Sony WF-1000XM3 earbuds by interfacing with the existing battery system, preserving all audio functionality while adding visual capabilities (Fig.~\ref{fig:blockdiagram}). We selected the Sony WF-1000XM3 platform based on accessibility for disassembly and battery integration, enabling a maintainable research scope without requiring a complete earbud redesign from the ground up. Modern wireless earbuds incorporate sophisticated components, including digital signal processors, multiple microphones, in-ear sensors, class D amplifiers, and custom application-specific integrated circuits (ASICs). By building upon an existing commercial platform, we demonstrate that camera integration is feasible within the constraints of commercial earbud designs rather than simplified prototypes.

Proceeding hardware integration, we designed custom 3D-printed enclosures that seal each modified earbud while positioning the camera sensor to achieve forward-facing capture aligned with the user's  field of view. The enclosure design maintains the overall form factor expectations of modern earbuds while accommodating the additional camera hardware. Camera positioning optimization and angular field of view considerations are detailed in~\xref{sec:modeling}.

\subsubsection{Camera and Bluetooth Chip Integration} 
Since the nRF52840 Bluetooth Low Energy SoC lacks a dedicated camera interface, we configure the HM01B0 CMOS sensor to operate in single-wire data transfer mode over its Display Video Port (DVP) interface. In this configuration, the HM01B0 effectively functions as an SPI controller, transmitting pixel data to the nRF52840 through its SPI peripheral (SPIS) port. Image transfer is coordinated through three key signals: Frame Valid (FVLD) indicates frame start with a rising edge, while Pixel Clock Out (PCLKO) and Data (D0) deliver synchronized pixel data. These signals map to the standard SPI protocol as chip select (CS), serial clock (SCLK), and controller-out-peripheral-in (MOSI). 

Clock synchronization between the camera sensor and microcontroller is achieved by generating an 8~MHz master clock (MCLK) signal from the nRF52840 to drive the HM01B0, ensuring the pixel clock output remains within the nRF52840's maximum SPI frequency specification of 8MHz. This approach maintains tight timing synchronization between both devices throughout the image capture process. The interface requires signal polarity adaptation since the nRF52840 SPI peripheral expects an active-low chip select signal, while the HM01B0's FVLD output is active-high. We address this by configuring FVLD as a GPIO interrupt source that triggers an external loopback connection, generating the required active-low chip select signal for the SPI interface. This approach could be optimized in future revisions by incorporating an inverter gate between FVLD and the SPI chip select line~ \cite{RoboticsPaper, iris}.
 Finally, the nRF52840's SPI DMA controller imposes a 64~kB maximum transfer size, while a full 324×239 pixel image requires 77.4~kB of data transfer. We overcome this limitation by splitting each image capture into two sequential SPI transactions, using the external GPIO loopback to trigger the second transaction mid-frame.

\subsubsection{Low-Power Optimizations}
The VueBuds camera module implements a three-state power management architecture to minimize energy consumption: OFF, IDLE, and ACTIVE. In the OFF state, the camera module is completely powered down, corresponding to scenarios where the earbuds are removed from the user's ears (detected via in-ear proximity sensors) or placed in the charging case. The IDLE state represents the standby mode, where VueBuds hardware is powered on with an interrupt from the earbud's primary SoC. In this state, the nRF52840 maintains an active Bluetooth Low Energy connection to the host device, while the HM01B0 camera  remains configured via I2C but clock-gated to minimize power consumption. The system transitions from IDLE to ACTIVE state upon wake-word detection (e.g., "Hey Vue" or "VueBuds"), enabling immediate camera activation without configuration delays. This wake-word paradigm mirrors existing earbud voice  interactions such as "Hey Siri" or "Ok Google," facilitating seamless adoption within established user interaction patterns.

During ACTIVE operation, VueBuds enables full camera functionality and streams visual data to the connected host device. To balance responsiveness with power efficiency, the system automatically returns to IDLE state after 3 seconds of streaming, ensuring minimal impact on overall earbud battery life while maintaining user interaction capabilities. This strategy enables VueBuds to operate with minimal impact on the existing earbud power budget. 

\begin{table*}[t!]
    \centering
    \caption{Design space for dual ear-level camera orientation.}  
    \Description{Table showing the effect of camera angular orientation on blind spot length, field of view, and overlap at Harmon distance. At 0°, the forward blind spot length is 14.1 cm with a stereo field of view of 88° and 64\% overlap. At 5° and 10°, the blind spot lengths increase to 18.6 cm and 24.7 cm, with stereo fields of view of 98° and 108°, and overlap decreasing to 46\% and 28\%, respectively. At 15°, the blind spot reaches 34.0 cm, with a stereo field of view of 118° and 14\% overlap. At 20°, the blind spot increases to 50.7 cm, expanding the stereo field of view to 128°, but no overlap remains at Harmon distance.}
    \vskip -0.15in
\begin{tabular}{ |c|c|c|c|c| }
\hline
Angular Orientation & Forward Blind Spot Length & Added Field of View & Binocular Field of View & Overlap at Harmon Distance\\
\hline
0\textdegree & 14.1cm & 0\textdegree & 88\textdegree & 64\% \\
\textbf{5\textdegree} & \textbf{18.6cm} & \textbf{10\textdegree} & \textbf{98\textdegree} & \textbf{46\%}\\
\textbf{10\textdegree} & \textbf{24.7cm} & \textbf{20\textdegree} & \textbf{108\textdegree} & \textbf{28\%} \\
15\textdegree & 34.0cm & 30\textdegree & 118\textdegree & 14\%\\
20\textdegree & 50.7cm & 40\textdegree & 128\textdegree & N/A \\
\hline
\end{tabular}
   \label{tab:blind-spot-modeling}
    \vskip -0.15in
\end{table*}

\subsubsection{Frame Rate and Wireless Latency} \label{subsec:wireless_latency}
 To maximize frame rate, we configure the system with the shortest connection interval permitted by the BLE specification (7.5ms) while transmitting 5 packets of 247 bytes per interval.  VueBuds operates at the maximum supported Bluetooth Low Energy data rate of 2 Mbps using LE 2M PHY~\cite{nordicsemi_bluetooth_range}. This configuration achieves approximately 992 kbps of effective throughput~\cite{maxthroughput}.

We evaluate two camera configurations from the HM01B0 sensor: QQVGA (162×119 pixels) and QVGA (324×239 pixels), generating 19.3 kB and 77.4 kB per frame, respectively. Given the maximum available BLE throughput, the theoretical wireless transmission limits are 6.4 fps for QQVGA and 1.6 fps for QVGA operation.
However, image acquisition introduces additional latency through the 8MHz SPI interface, requiring 19.3ms for QQVGA capture and 77.4ms for QVGA capture. Accounting for both acquisition and transmission overhead, the system achieves effective frame rates of 5.7 fps and 1.4 fps for QQVGA and QVGA configurations, respectively (see Table.~\ref{tab:latencytradeoff}). These rates could be improved through pipelining  that overlaps image acquisition with wireless transmission~\cite{pipelining}.

\begin{table}[h!]
    \centering
    \vskip -0.1in
    \caption{{\bf Resolution against frame rate and latency.} }
    \Description{Table showing the latency and frame rate tradeoff for VueBuds camera configurations. At QQVGA resolution (162×119), the system achieves 5.7 frames per second with a frame latency of 175 ms. At QVGA resolution (324×239), the frame rate decreases to 1.4 frames per second with a frame latency of 714 ms.}
    {
    \vskip -0.15in
    \begin{tabular}{ |c|c|c| }
        \hline
        Camera Configuration & Frame Rate & Frame Latency \\
        \hline
        QQVGA (162×119) & 5.7 fps & 175 ms \\
        QVGA (324×239) & 1.4 fps & 714 ms \\
        \hline
    \end{tabular}
    \label{tab:latencytradeoff}
    }
   \vskip -0.15in
\end{table}


\subsubsection{Fabrication}
We integrate custom 3D-printed enclosures with commercial Sony WF-1000XM3 earbuds to house the camera hardware. The enclosures were designed in OnShape CAD and manufactured on a Bambu Labs A1 Mini 3D printer. VueBuds printed circuit boards were designed with the open-source KiCad eCAD tool and fabricated as 2-layer PCBs through PCBWay, with component assembly completed by a local  contractor. The enclosure design ensures seamless integration with the original earbud form factor while providing precise camera positioning. The HM01B0 camera sensor is mounted behind a window in the enclosure lid, oriented forward to capture the user's natural field of view. This positioning aligns the camera with the user's perspective when the earbuds are worn, facilitating intuitive visual queries by allowing users to simply look toward objects of interest.

\subsection{Binocular Vision for Ear-Level Cameras}
\label{sec:blind-spot-method}

\subsubsection{Camera Field-of-View and Facial Occlusion}
We first quantify the extent of facial occlusion introduced by ear-level positioning and demonstrate how sensor windowing reduces obstruction while introducing coverage trade-offs. The HM01B0 camera sensor provides an 87-degree horizontal field of view, which poses significant challenges when positioned at ear level on wireless earbuds. Unlike temple-mounted cameras on smart glasses, which benefit from forward positioning relative to the user's face, earbud-mounted cameras suffer from substantial facial obstruction, with the user's face blocking a considerable portion of each camera's field of view, specifically in the inward-facing regions.

Facial obstruction severely impacts the ability to capture egocentric viewpoints and can cause issues with vision language model interpretation and panoramic stitching correspondence. To mitigate this obstruction, we implement the HM01B0's windowed readout mode, which crops the sensor's active area to reduce the effective field of view from 87° to 65°. This windowing approach eliminates the facial obstruction zone while (1) maintaining forward-facing capture capability, and (2) reducing unnecessary data transmission over the bandwidth-constrained Bluetooth Low Energy link.


This 25\% reduction in captured visual data creates a fundamental trade-off: while windowed readout helps resolve facial obstruction, it significantly reduces peripheral visual coverage, particularly limiting the left and right outskirts of the user's natural field of view. This necessitates alternative strategies to recover comprehensive environmental coverage through dual-camera coordination.

\begin{figure*}[t!]
\centering
\includegraphics[width=1\linewidth]{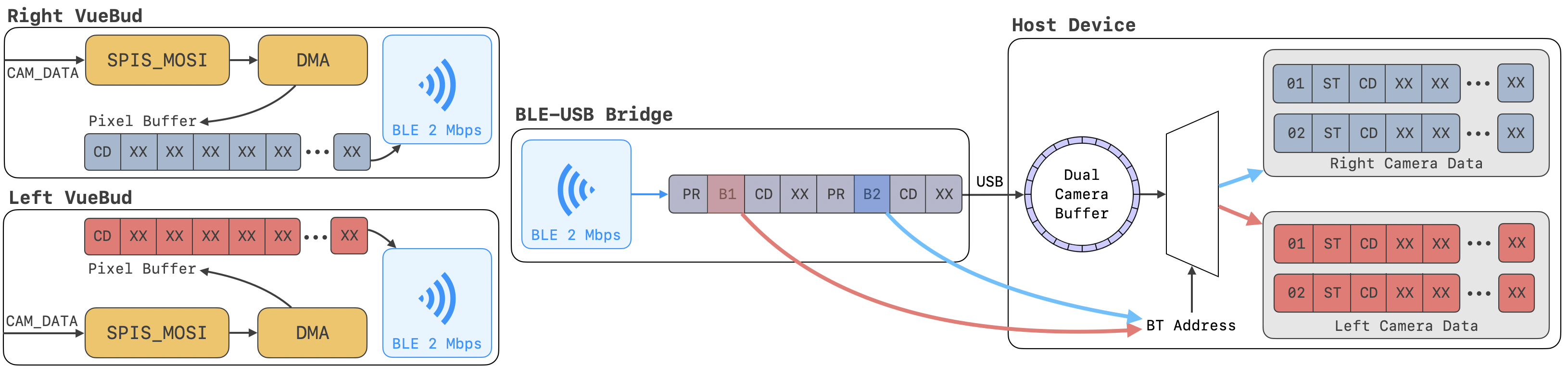}
\vskip -0.15in
\caption{{\bf VueBuds to host device wireless data streaming.} \textmd{ VueBuds stream dual camera data to the BLE-USB bridge, which multiplexes the binocular image feed for transmission to the host device, where they are ultimately demultiplexed into left and right images.}}
\Description{Diagram of the camera-to-host device data pipeline for VueBuds. The left and right cameras capture pixel data through the SPIS_MOSI interface, which is buffered and transferred via DMA before being streamed over Bluetooth Low Energy at 2 Mbps. A BLE-to-USB bridge collects the packetized data, appends preambles and device addresses, and forwards it to the host device. On the host, dual camera buffers separate left and right camera streams based on device addresses for frame reconstruction and visual language model input.}
\vskip -0.15in
\label{fig:integration}
\end{figure*}

\subsubsection{Camera Angular Orientation and Blind Spot Characterization} \label{sec:modeling}

 To compensate for the visual context lost through windowed readout, we systematically evaluate camera angular orientations at 0°, 5°, 10°, 15°, and 20° outward from the forward-facing position. This angular adjustment serves the dual purpose of reducing potential facial obstructions and expanding the combined field of view.

In the completely forward-facing orientation (0°), windowed readout enables VueBuds to capture approximately 88° of the user's forward perspective per camera. However, outward camera angling introduces competing design constraints that must be carefully balanced. Angling cameras outward conflicts with our core design principle of capturing the user's natural forward-facing perspective and creates an expanding ``blind spot'' directly in front of the user, where objects held at close distances fall outside both cameras' fields of view.

We address this trade-off by incorporating the Harmon distance, the average comfortable reading distance of 36.8cm established across 233 individuals~\cite{viewingDistSmartphone}. Objects held closer than this distance typically fall outside normal interaction patterns for reading and detailed visual examination. By designing around this ergonomic constraint, we establish a tolerable blind spot that minimally impacts practical usage scenarios.

To characterize the forward-facing blind spot region, we model the binocular camera field of view on a person's head at 5-degree increments until the forward blind spot length exceeds the Harmon distance. We define this blind spot region as the distance from a person's eye to where a 5 cm wide object must be present in each camera's field of view to ensure sufficient correspondence for image stitching algorithms and provide conservative redundancy between left and right images. Our analysis in Table.~\ref{tab:blind-spot-modeling} demonstrates that camera orientations of 5° and 10° maintain forward blind spot lengths well below the Harmon distance (18.6~cm and 24.7~cm, respectively), while adding 10-20° of additional field of view. At 15°, the blind spot (34.0 cm) approaches the Harmon distance threshold, representing a practical limit for maintaining usable interaction patterns. Beyond 15° of outward angling, the blind spot expansion significantly degrades user experience, potentially requiring users to step backward or extend objects to uncomfortable distances when looking directly at an object. We verify our blind spot simulations using our hardware in ~\xref{sec:blind-spot-eval}.

\subsubsection{Image Stitching for Redundancy Reduction} While the simple approach is to feed each camera image from the left and right earbuds into the vision language model, we also explore image stitching to reduce redundant visual information and improve vision language model inference runtime.

Overlaps exist between left and right image pairs captured by VueBuds, particularly in the far-field regions. At the Harmon distance, we calculate 64\%, 46\%, and 28\% overlap between images at 0°, 5°, and 10° camera orientations, respectively. To reduce effective input tokens and improve end-to-end latency, we implement lightweight stitching using ORB (Oriented FAST and Rotated BRIEF) feature detection, selected for computational efficiency~\cite{6126544}. {At a high level, these techniques work by detecting repeatable keypoints in both images,  converting them to compact binary feature vectors,  and then matching corresponding features to estimate geometric transformations for alignment and stitching.} Our stitching pipeline operates without post-processing operations such as image trimming to preserve maximum visual information for the vision language model and minimize processing latency. 



\subsection{End-to-End System Integration}
Shown in Fig.~\ref{fig:integration}, we demonstrate wireless data flow from VueBuds hardware to vision language model processing.

\subsubsection{BLE-USB Bridge}
The system employs an nRF52840 development board as a BLE-USB bridge that converts Bluetooth Low Energy communication to USB connectivity for the host processing device. The receiver establishes connections with both VueBuds devices and manages both camera data streams simultaneously.

Upon receiving packets from each earbud, the receiver prepends a fixed preamble and the originating device's Bluetooth MAC address to enable stream demultiplexing. This packetized data is transmitted via USB serial to the host device, where our Python-based parsing module processes the incoming stream. The parser uses the preamble for packet boundary detection, the MAC address to route data into separate left and right image buffers, and a frame indicator to detect image boundaries for frame reconstruction.

\subsubsection{Vision Language Model Processing}
We implement on-device vision language model processing using Ollama, evaluating five models for optimal performance characteristics: Qwen2.5VL~\cite{bai2025qwen25vltechnicalreport}, Moondream~\cite{moondream}, MiniCPM-V~\cite{yao2024minicpmvgpt4vlevelmllm}, LLaVA~\cite{liu2023visualinstructiontuning}, and Gemma3~\cite{gemma3}. Model selection is based on systematic evaluation across accuracy and latency metrics, detailed in \xref{sec:vlmresults}.  Our implementation incorporates an adaptive system prompt that adjusts based on image input type, providing context-aware prompts for stitched images versus dual independent camera feeds. While the system supports both 162×119 and 324×239 resolutions, we select the higher resolution configuration based on its superior text recognition performance observed during evaluation.

\subsubsection{Audio Processing Pipeline}
Our audio pipeline uses the Bluetooth Hands-Free Profile for audio data transfer, and incorporates automatic speech recognition (ASR) and synthesis (TTS) capabilities. Our system integrates TinyWhisper for real-time speech-to-text conversion, chosen for its computational efficiency and accuracy in wake-word detection.

Our wake-word detection implementation uses a fuzzy matching algorithm to handle variations in pronunciation and ambient noise levels. The system recognizes wake phrases including "VueBuds", "Hey Vue", and phonetic variations, triggering camera activation and visual query processing. The system incorporates confidence thresholding to minimize false-positive wake-word activations while maintaining responsiveness. Speech synthesis utilizes Apple's built-in text-to-speech engine to deliver VLM responses through the earbuds' integrated speakers, enabling private audio playback. The entire audio pipeline operates in tandem with the visual pipeline, using a multi-process architecture to maximize system performance.

\section{Evaluation}
{We present the system's vision capabilities along with accuracy and latency results across several vision language models. Then, we select the best-performing VLM for our system and share the results from our online and in-person user studies. Finally, we assess system performance in terms of power and end-to-end latency.}

\begin{figure*}[t!]
\vskip -0.1in
\centering
\includegraphics[width=0.9\linewidth]{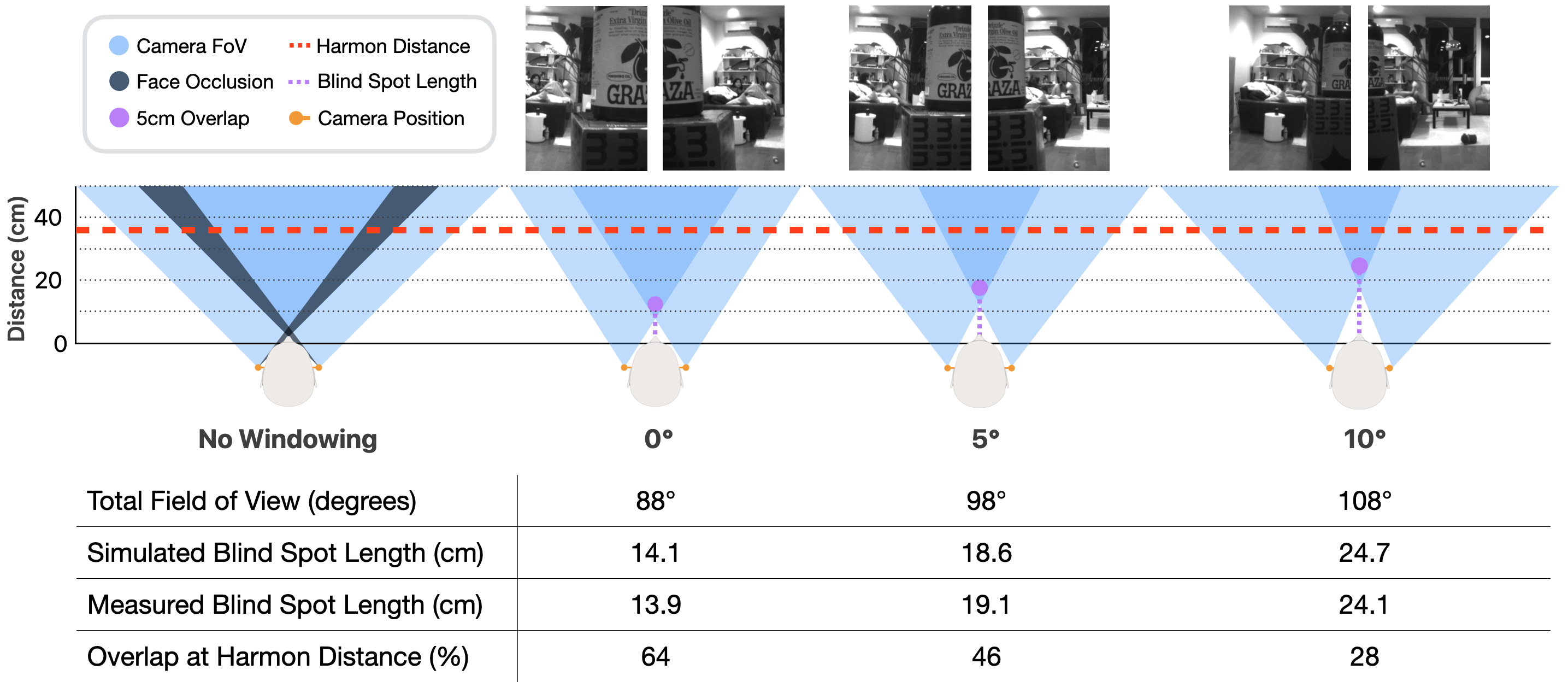}
\vskip -0.15in
\caption{{\bf Blind spot evaluation.} \textmd{Geometric modeling of dual camera field of view (blue) showing facial occlusion and blind spot regions at various angular orientations. The red line indicates the Harmon distance. Empirical measurements (top) closely align with the model.}}
\Description{Diagram showing the blind spot evaluation for VueBuds at different camera orientations. The top row shows images captured at 0°, 5°, and 10° outward angles, alongside a baseline with no windowing. The visualized fields of view illustrate how blind spots expand as cameras are angled outward. The table below reports simulated and measured blind spot lengths, with simulated values of 14.1 cm, 18.6 cm, and 24.7 cm at 0°, 5°, and 10° respectively, and measured values of 13.9 cm, 19.1 cm, and 24.1 cm. The total field of view expands from 88° with no windowing to 108° at 10°, while overlap at Harmon distance decreases from 64\% at 0° to 28\% at 10°.}
\vskip -0.15in
\label{fig:blind-spot}
\end{figure*}

\subsection {Vision Capability Benchmarks}
Here, we evaluate VueBuds' tolerance to blind spots within our camera angle design space, and then present VLM accuracy and latency results using images from our camera to explain why we chose Qwen2.5VL 7B among 5 selected vision language models.

\subsubsection {Blind Spot Evaluation}\label{sec:blind-spot-eval}
As discussed in \xref{sec:blind-spot-method}, the usability of camera-integrated earbuds depends critically on capturing the user's forward field of view while preserving natural interaction patterns. Users should not need to hold objects at awkward angles or step backward to accommodate system limitations.  

\textbf{Methodology.} In \xref{sec:modeling}, we developed a geometric model to project the dual camera fields of view from ear-mounted positions. Using a mannequin wearing VueBuds, we measured a lateral camera offset of 2.3 cm. Our model projects the windowed 65° field of view from each camera position at angular orientations of 0°, 5°, 10°, 15°, and 20° outward from the forward-facing direction. The blind spot length was defined as the distance from the user's eye center to the point where a 5 cm object would be captured in its entirety by both cameras' fields of view. 

 To validate our theoretical values, we constructed an adjustable test rig using L-brackets that allowed precise angular positioning of the VueBuds cameras. For each angular configuration, we empirically measured the blind spot using a circular bottle with a text label as our test object. We placed the bottle at the mannequin's eye center and gradually moved it backward until visual commonality was confirmed across both left and right camera views. This methodology directly measured the practical blind spot distance with our actual hardware system. As shown in Fig.~\ref{fig:blind-spot}, our empirical measurements closely aligned with the geometric model across all tested configurations. This strong correspondence (within 3\% error) validated our geometric modeling approach for system design.


\textbf{Design Tradeoffs}. The total effective field of view, measured from the leftmost edge of the left camera to the rightmost edge of the right camera, expands from 88° at 0° orientation to 128° at 20°. As discussed in \xref{sec:modeling}, this expanded coverage comes at the expense of increasing the blind spot length. This adversely affects usability, as orientations beyond 15° potentially require users to hold objects farther than normal interaction distances. At 15°, the blind spot length (34.3 cm) approaches the Harmon distance threshold of 36.8 cm, while the 20° configuration creates a 49.5 cm blind spot that would severely degrade natural interaction patterns. Based on this analysis, we designed VueBuds with angular orientations between 5\textdegree and 10\textdegree, balancing expanded field of view coverage while preserving intuitive interaction where users can simply look toward objects of interest without adjusting their behavior.

\subsubsection{VLM Accuracy and Latency Performance}\label{sec:vlmresults}
Selecting an appropriate vision language model for VueBuds requires balancing accuracy, latency, and on-device deployment constraints. 

\textbf{Methodology.} We evaluated five models under 8B parameters: Qwen2.5-VL (7B), Moondream (1.8B), MiniCPM-V (8B), LLaVA (7B), and Gemma3 (4B); all supported through Ollama. Each model processed 20 scenes captured by VueBuds at both 160px and 320px resolutions across three task categories: (1) Object/Scene Recognition: identifying common objects and describing scenes, (2) Optical Character Recognition: reading text from signs, labels, and documents, and (3) Translation/Reasoning: answering questions requiring visual understanding and logical inference. All benchmarks were performed on a Mac Mini M4 Pro in its base configuration ~\cite{macminichoose}, with human evaluators verifying response accuracy.

\begin{figure*}[t!]
\centering
\includegraphics[width=1.0\linewidth]{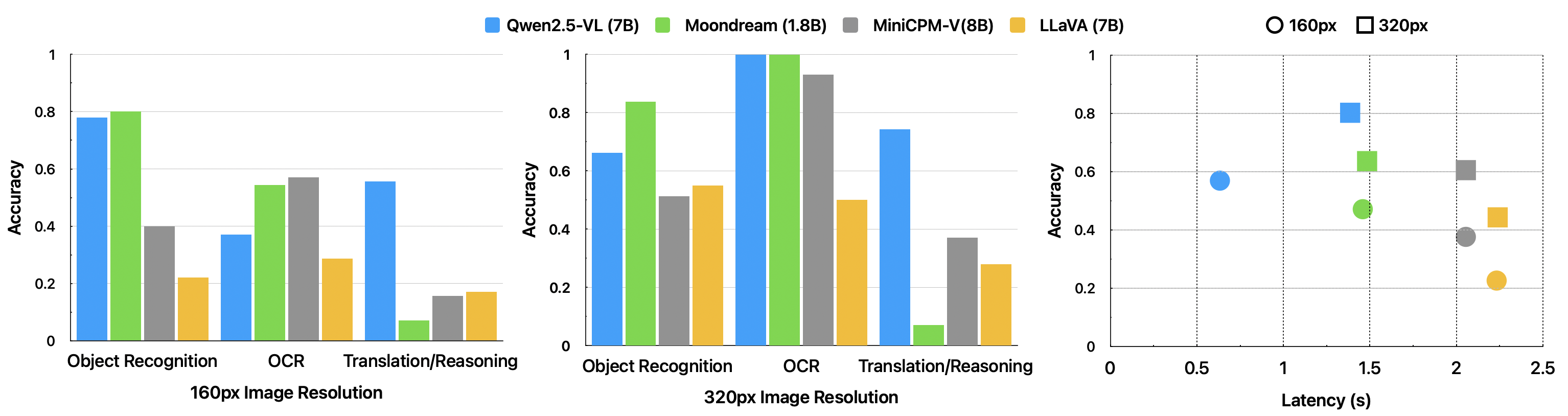}
\vskip -0.1in
\caption{{\bf VLM comparison.} \textmd  {Task-specific accuracy at 160px (left) and 320px (center), and accuracy against latency (right).} 
\Description{This figure contains three plots showing VLM performance across different resolutions and latency trade-offs. The left bar chart shows task-specific accuracies at 160px image resolution, with models compared on object recognition, OCR, and translation tasks.The middle bar chart shows the same tasks evaluated at 320px resolution, where all models improve substantially, particularly for OCR accuracy. The right scatter plot shows the trade-off between accuracy and latency across the evaluated models, illustrating how higher accuracy often comes at the cost of increased response time.}}
\Description{}
\vskip -0.15in
\label{fig:vlm_results}
\end{figure*}

\textbf{Resolution Impact.} Fig.~\ref{fig:vlm_results} shows that increasing resolution from 160px to 320px (a 4x increase in total pixels) yields substantial accuracy improvements across all models. Overall accuracy increased by 35\% for both Qwen2.5-VL and Moondream, 61\% for MiniCPM-V, and 96\% for LLaVA. The impact was most pronounced for text reading tasks, where Qwen2.5-VL and Moondream achieved perfect accuracy at 320px within our dataset, representing improvements of 170\% and 84\% respectively from their 160px baselines. This suggests OCR capabilities degrade rapidly once character boundaries become indistinguishable at lower resolutions. For queries like "What does this sign say?" or "Can you translate this for me?", the higher resolution proves essential. This validated our decision to use HM01B0's 324x239 resolution despite the increased  latency.

\textbf{Accuracy-Latency Tradeoffs.} Response latency directly impacts user experience in conversational interfaces. Fig.~\ref{fig:vlm_results} illustrates this design space, with time-to-first-token ranging from 0.6 seconds (Qwen2.5-VL at 160px) to 2.24 seconds (LLaVA at 320px). Notably, Gemma3 exhibited severe latency issues, averaging over 12 seconds to first token, a known KV cache quantization bug~\cite{ollama2024gemma2} that rendered it challenging for real-time application use  despite its compact 4B parameter size.  Qwen2.5-VL achieved the optimal balance, delivering the highest overall accuracy (80.1\%) while maintaining the lowest latency (1.39s at 320px). Its architectural optimizations, including window attention in the vision transformer~\cite{qwen2025vl}, enable efficient processing along with its dynamic resolution adaptation. While Qwen2.5-VL shows the largest relative latency increase when scaling from 160px to 320px, this corresponds to its dynamic resolution processing that adjusts the number of visual tokens based on input complexity. Crucially, even at 320px, Qwen2.5-VL remains the fastest model in our evaluation.

\textbf{Model Selection Rationale.} While matching Qwen2.5-VL on object recognition and OCR tasks, Moondream failed on reasoning queries (7.1\% accuracy), often returning empty responses for complex questions~\cite{gitollama, gitollama2}. This limitation disqualifies it for queries requiring inference, such as "How many calories are in this?" and similar questions. MiniCPM-V exhibited overfitting to our system prompts, generating templated responses that failed to adapt to varied user inputs. LLaVA produced inconsistent outputs when processing our low-resolution monochrome images. These evaluations confirm Qwen2.5-VL as a better model for VueBuds, providing reliable performance across diverse visual queries while maintaining sub-1.5-second response times. 

\subsection{User Studies} \label{sec:studies}

{To comprehensively evaluate VueBuds, we conducted user studies with 90 participants examining three components: platform accessibility, response quality compared to Ray-Ban Meta, and real-world performance across standardized visual question answering (VQA) tasks. The first two components were administered as a two-part online survey through Google Forms to efficiently gather comparative assessments at scale, while the third was conducted through in-person sessions to evaluate real-world performance. Each study used a within-subject design with no compensation. All participants were informed of the study's purpose and procedures and voluntarily agreed to participate. User data was anonymized, and no photos containing participants' faces were retained. All studies were approved by our university's Institutional Review Board.}

\begin{figure*}[t!]
\vskip -0.1in
\centering
\includegraphics[width=1.0\linewidth]{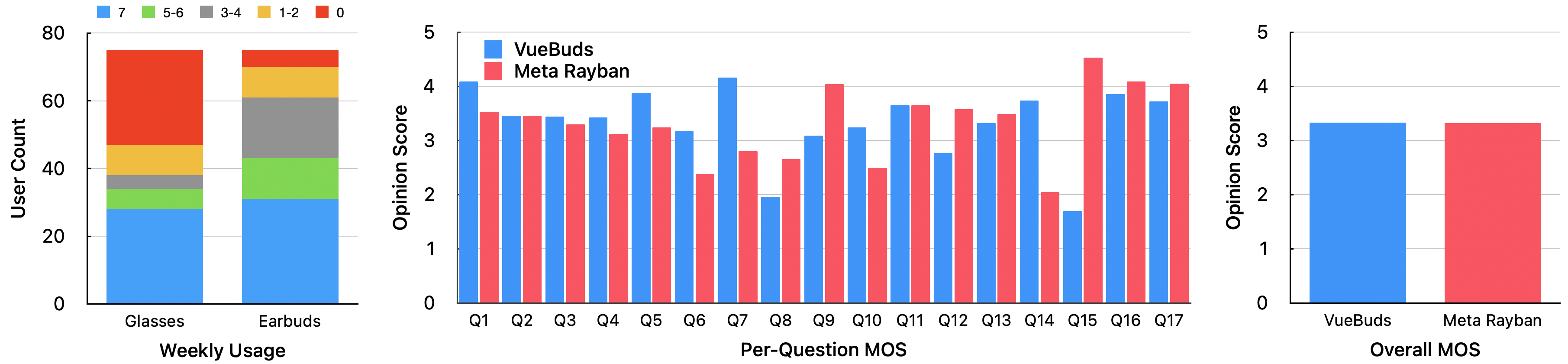}
\vskip -0.15in
\caption{{\bf Online user study results.}  \textmd {(Left) Weekly usage patterns for regular (non-smart) glasses and earbuds.  (Middle) Mean Opinion Score (MOS) of VueBuds against Ray-Ban Meta for each of the 17 visual question answering tasks. (Right) MOS averaged across all 17 tasks.}}
\Description{The figure shows three sub-figure comparing the user study results for VueBuds and Ray-Ban Meta smart glasses. The left sub-figure  is a  stacked bar graph titled "Weekly Usage". It shows the weekly usage patterns for regular (non-smart) glasses and earbuds. The Y-axis represents User Count, ranging from 0 to 80. The X-axis has two bars: "Glasses" and "Earbuds".
The bars are segmented by color, representing different weekly usage frequencies. The middle sub-figure shows the Per-Question Mean Opinion Score (MOS). It is a  grouped bar graph titled "Per-Question MOS". It compares the Mean Opinion Score (MOS) for VueBuds and Ray-Ban Meta across 17 visual-language question-answering tasks (Q1 to Q17). The Y-axis represents the Opinion Score, ranging from 0 to 5. The X-axis lists the questions from Q1 to Q17. Blue bars represent VueBuds. Red/Pink bars represent Ray-Ban Meta. The right sub-figure  is a simple bar graph titled "Overall MOS". It shows the average Mean Opinion Score across all 17 tasks. The Y-axis represents the Opinion Score, ranging from 0 to 5. The X-axis has two bars: "VueBuds" and "Ray-Ban Meta". The VueBuds bar (blue) and the Ray-Ban Meta bar (red/pink) are nearly identical in height. This indicates that the overall user opinion for both devices was very similar.}
\vskip -0.05in
\label{fig:survey-mos}
\end{figure*}

\begin{figure*}
\centering
\includegraphics[width=0.99\linewidth]{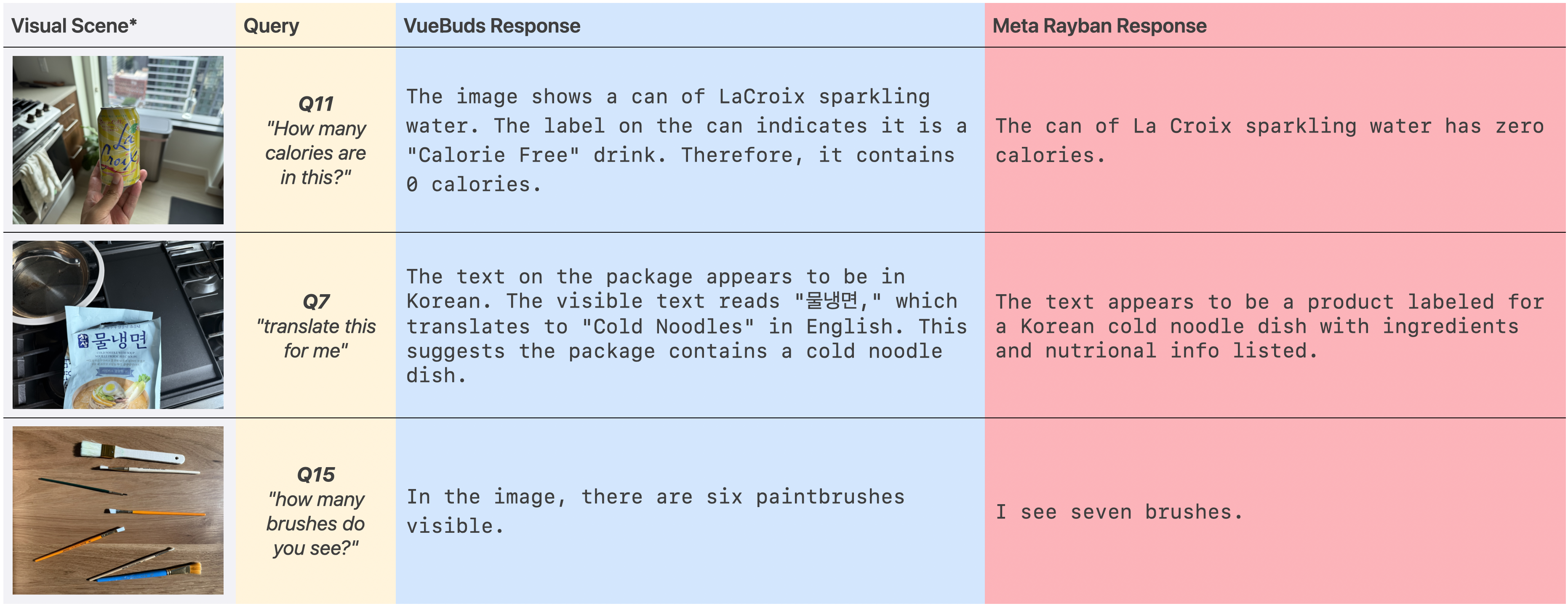}
\vskip -0.1in
\caption{{\bf Samples from the online user study.} \textmd{Visual scenes(*) presented to participants were captured using a mobile phone to eliminate camera quality bias when comparing response quality between VueBuds and Ray-Ban Meta.}}
\Description{The figure presents three examples of a user's query and the corresponding responses generated by two different smart glasses: VueBuds and Ray-Ban Meta. The table has four columns: Visual Scene: This column contains an image of the scene. Query: The question posed by the user. VueBuds Response: The answer provided by the VueBuds. Ray-Ban Meta Response: The answer provided by the Ray-Ban Meta smart glasses.

Example 1: Calorie Count
Visual Scene: A person holding a can of LaCroix sparkling water.
Query: "How many calories are in this?" (Q11)
VueBuds Response: "The image shows a can of LaCroix sparkling water. The label on the can indicates it is a 'Calorie Free' drink. Therefore, it contains 0 calories.
Ray-Ban Meta Response: "The can of La Croix sparkling water has zero calories."

Example 2: Translation
Visual Scene: A package with Korean text on it.
Query: "Translate this for me." (Q7)
VueBuds Response: "The text on the package appears to be in Korean. The visible text reads '물냉면,' which translates to 'Cold Noodles' in English. This suggests the package contains a cold noodle dish."
Ray-Ban Meta Response: "The text appears to be a product labeled for a Korean cold noodle dish with ingredients and nutritional info listed."

Example 3: Object Counting
Visual Scene: A view of several paintbrushes lying on a table.
Query: "How many brushes do you see?" (Q15)
VueBuds Response: "In the image, there are six paintbrushes visible."
Ray-Ban Meta Response: "I see seven brushes."}
\vskip -0.15in
\label{fig:user-study-highlights}
\end{figure*}

\subsubsection{Study 1: Platform Accessibility Analysis} \label{sec:platform-accessibility}

Here, we examine device usage patterns to establish the potential user base for earbud-based and glasses-based wearables, providing critical context for earbud accessibility and adoption compared to glasses platforms.

{\textbf{Participants.} For the online portion of our study, we recruited 74 participants (48 male, 26 female) between the ages of 19--65 ($\bar{x}=35.9$, $\sigma=13.93$) through convenience sampling via personal networks. Participants represented diverse professional backgrounds, including engineers, healthcare professionals, designers, consultants, marketers, students, stay-at-home parents, and retirees. The majority of participants were located in the United States, with additional participants from South Korea, Hong Kong, and Japan. }

{\textbf{Methodology.}} We conducted an online survey examining participants' usage frequency of regular (non-smart) glasses and earbuds. Participants answered two questions about their device usage frequency: \textit{1) How often do you wear (ordinary) glasses?}, and \textit{2) How often do you wear earbuds or headphones?}. Response options included: Everyday, 5-6 days a week, 3-4 days a week, 1-2 days a week, and Never. We then categorized responses into three usage patterns: high-frequency users (5-7 days per week), occasional users (1-4 days per week), and non-users (never). Chi-square tests compared usage pattern distributions between glasses and earbuds. 

{\textbf{Results.}} Analyzing the usage patterns revealed striking differences in device adoption. For glasses, 45.3\% were high-frequency users, 17.3\% occasional users, and 37.3\% non-users. In contrast, earbuds showed substantially higher engagement: 57.3\% high-frequency users, 36\% occasional users, and only 6.7\% non-users. This translates to 93.3\% of participants using earbuds at least occasionally, compared to 62.7\% for glasses.

A chi-square test confirmed that usage patterns differed significantly between device types ($\chi^2(2) = 21.98, p < 0.001$), with the largest differences in non-user rates. The exclusion rates provide the most compelling insight: over one-third of participants (37.3\%) never wear glasses, thereby immediately excluding them from any glasses-based AI system. Conversely, only 6.7\% avoid earbuds entirely, suggesting that earbud-based platforms could reach a significantly larger user base.

These findings indicate that camera-integrated earbuds like VueBuds could reach a broader audience than glasses-based systems, particularly among the 37.3\% of users who do not wear glasses. While this data reflects comfort with traditional device categories rather than AI acceptance, it establishes that earbud-based systems face fewer baseline accessibility barriers, potentially reducing adoption friction for wearable AI technologies.

\begin{figure*}[t!]
\vskip -0.1in
\centering
\includegraphics[width=0.95\linewidth]{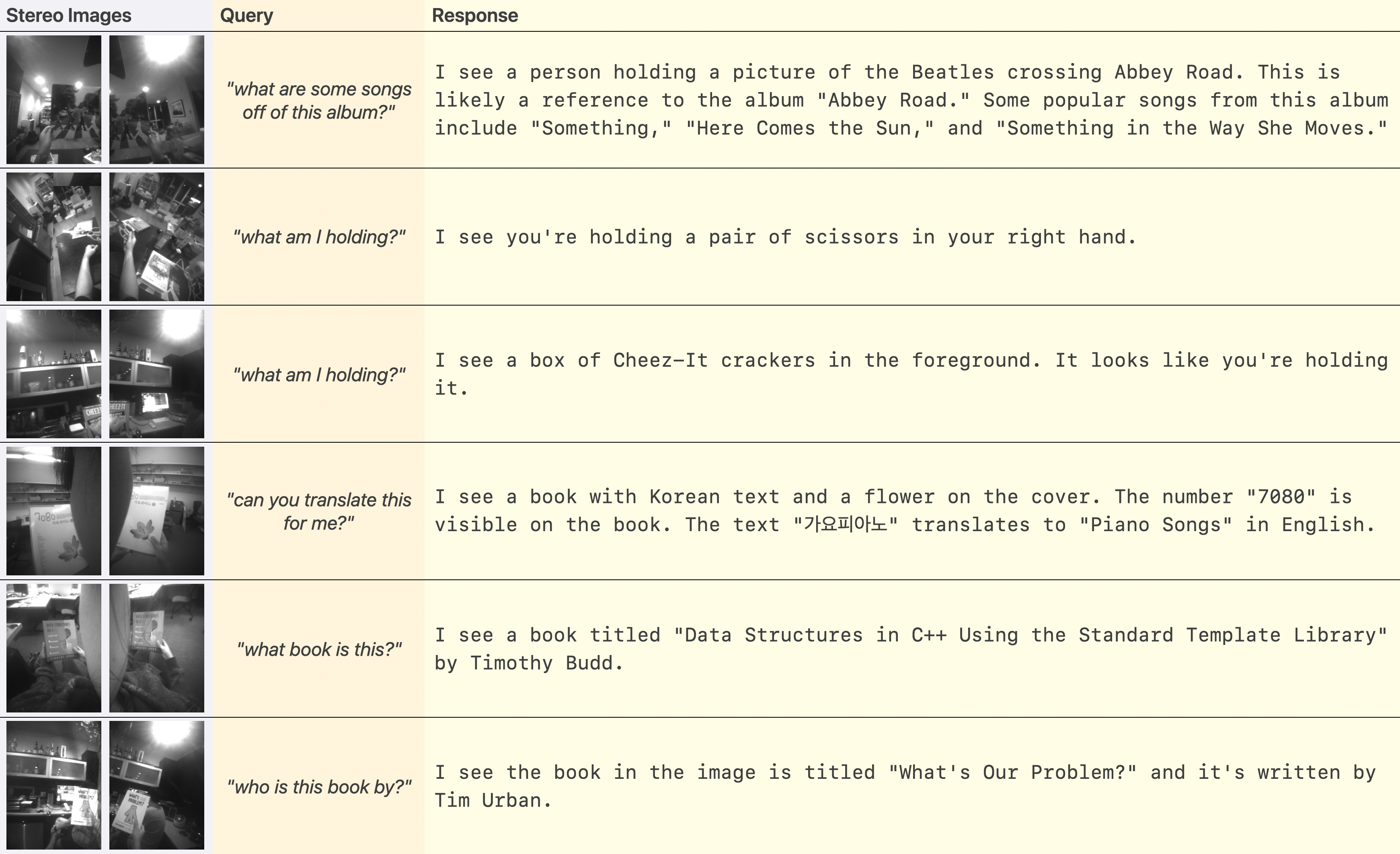}
\vskip -0.1in
\caption{{\bf VueBud query and response samples from the in-person user study.} \textmd {VueBuds captures the binocular imagery (left) and user query (middle) which is then provided as input to the vision language model for an auditory response (right).}}
\Description{Image contains grid of stereo images paired with example queries and responses from VueBuds during in-person user studies. The leftmost column shows grayscale images captured from the earbuds. The middle column contains user queries such as “what am I holding?” or “can you translate this for me?”. The right column shows VueBuds’ responses. Example outputs include identifying scissors, a box of Cheez-It crackers, translating Korean text as “Piano Songs,” recognizing the book Data Structures in C++ Using the Standard Template Library by Timothy Budd, and identifying the book What’s Our Problem? by Tim Urban. The first row also shows album cover recognition, correctly identifying the Beatles’ Abbey Road and listing songs from it.}
\vskip -0.15in
\label{fig:user-study-highlights}
\end{figure*}

\subsubsection{Study 2: Response Quality Evaluation Against Ray-Ban Meta} \label{sec:vs-rayban}
Here, we evaluate the subjective response quality between VueBuds and commercial smart glasses.

{\textbf{Participants.}  74 participants from Study 1 completed this study.}

{\textbf{Methodology.}} Participants performed a blind comparative evaluation of VueBuds and Ray-Ban Meta response quality across 17 visual question answering tasks using a 5-point Mean Opinion Score (MOS) scale. We designed 17 evaluation scenarios representing typical smart glasses use cases: scene recognition ("what do you see"), text recognition (book titles), translation, calorie estimation, plant care advice, and object counting. For each scenario, we collected responses from both systems viewing identical real-world scenes. Ray-Ban Meta responses were generated by pointing the device at the scene, issuing verbal queries, and transcribing the audio. VueBuds responses were generated by capturing the scene with our camera system and processing through Qwen2.5-VL 7B. 

To ensure fair comparison, we implemented several controls: (1) For questions 1--15, we showed the online participants device-agnostic images captured using a mobile phone pointed at the same scenes, then paired these images with both systems' responses to eliminate camera quality bias, and (2) randomized response order to prevent position bias. Questions 16--17 presented both device-agnostic images and images captured directly by VueBuds, alongside both systems' responses, allowing participants to compare response quality while also evaluating VueBuds' image quality.

Participants received the instruction: \textit{"Imagine you are wearing AI-enabled smart glasses. You will see images with sample questions and two different AI responses."} Responses were rated on a 1--5 scale: 1=Poor (inaccurate, confusing, unhelpful), 2=Fair (somewhat relevant but incomplete), 3=Good (reasonably accurate and helpful), 4=Very Good (accurate, clear, useful), and 5=Excellent (highly accurate, very clear, very helpful). Participants were instructed that response length should not influence ratings. {After completing the evaluation, participants were given the option to provide open-ended feedback (see Appendix) to elaborate on their rating criteria and explain the factors that drove high and low scores.}

{\textbf{Results.}} The overall MOS scores in Fig.~\ref{fig:survey-mos}  demonstrate near-statistical parity: VueBuds achieved 3.33 compared to Ray-Ban Meta's 3.32. However, per-question analysis revealed task-dependent variations. The largest performance gap favored VueBuds in a translation task (Q7: 4.1 vs 2.8), where users strongly preferred specific translations over Ray-Ban Meta's general scene description. Conversely, the largest gap favoring Ray-Ban Meta occurred in object counting tasks (Q15: 4.5 vs 1.7), revealing limitations in Qwen2.5VL's numerical accuracy. Most other tasks showed comparable performance, with differences typically under 0.5 points.

The near-identical overall scores suggest that despite variation in individual tasks, both systems provide comparable user experience. This demonstrates that low-resolution monochrome images from VueBuds, when processed by Qwen2.5VL, can achieve response quality comparable to commercial smart glasses.

{Analysis of participant feedback revealed five primary criteria driving quality ratings: (1) \textit{Accuracy} - factual correctness, with counting errors particularly penalized; (2) \textit{Directness vs. verbosity} - concise, direct answers preferred over unnecessary elaboration; (3) \textit{Task-appropriate responses} - context-dependent quality expectations, such as actual translations for translation tasks; (4) \textit{Conversational suitability} - responses similar to everyday conversation; and (5) \textit{Confidence calibration} - appropriate uncertainty expression valued over overconfident incorrect responses.}

\subsubsection{Study 3: In-Person Performance Evaluation with VueBuds} \label{sec:inpersonstudy} Here, we show the results from an in-person study to evaluate VueBuds under realistic usage conditions, including variation in head shapes, earbud positioning, object holding distances, lighting conditions, and potential visual occlusions (e.g., glare, shadows).


{\textbf{Participants.}{ We recruited 16 participants (10 male, 6 female) aged of 20--44 ($\bar{x}=27.9$, $\sigma=6.9$) locally in the United States, including students (7), healthcare professionals (3), engineers (3), consultants (2), and a professor (1).}}

{\textbf{Methodology.}} Participants were tested individually across kitchen, office, and living room environments. Each participant was provided with VueBuds and informed that these were camera-integrated earbuds designed for visual question answering (VQA). After briefing participants on VueBuds' forward-facing camera design, we provided diverse test objects including snack boxes, food packages, kitchen utensils, books, vinyl records, and Korean-language items (signage, piano books, snack packaging). 

Participants were instructed to handle objects naturally while asking standardized questions to ensure consistent task categorization. To manage the open-ended nature of VLM capabilities, we focused evaluation on three primary tasks: Object Recognition, Optical Character Recognition (OCR), and End-to-End Translation. {These tasks were chosen as they represent foundational and critical real-world utilities (situational awareness \cite{htike2020ability, SystemDesign2025, SmartVisionGlasses2025, 10449230}, text access \cite{khan2020ai, baig2024ai, SmartVisionGlasses2025}, and multilingual communication \cite{SystemDesign2025, 10449230}) required for current and next-generation wearable vision assistance systems}:

\squishlist
\item{\textit{Object Recognition}: "What am I holding?" tasks testing basic visual identification}
\item{\textit{Optical Character Recognition (OCR)}: Book title and author identification requiring text extraction}
\item{\textit{End-to-End Translation}: Korean-to-English translation combining OCR and linguistic processing}
\squishend

All system responses were recorded in real-time and the responses were evaluated by comparing with the ground truth visual scene. Object recognition tasks received binary scoring (correct/incorrect), while OCR accuracy was measured using word error rate (WER). Translation tasks employed a two-stage evaluation: OCR accuracy for text extraction, followed by translation quality given successful text recognition. This approach isolated whether translation failures stemmed from visual processing or linguistic capabilities.  {After the study, we sent the same 16 participants a short survey (see Appendix) to provide additional qualitative insight regarding their experience with VueBuds.}

\begin{figure}[t!]
\centering
\includegraphics[width=\linewidth]{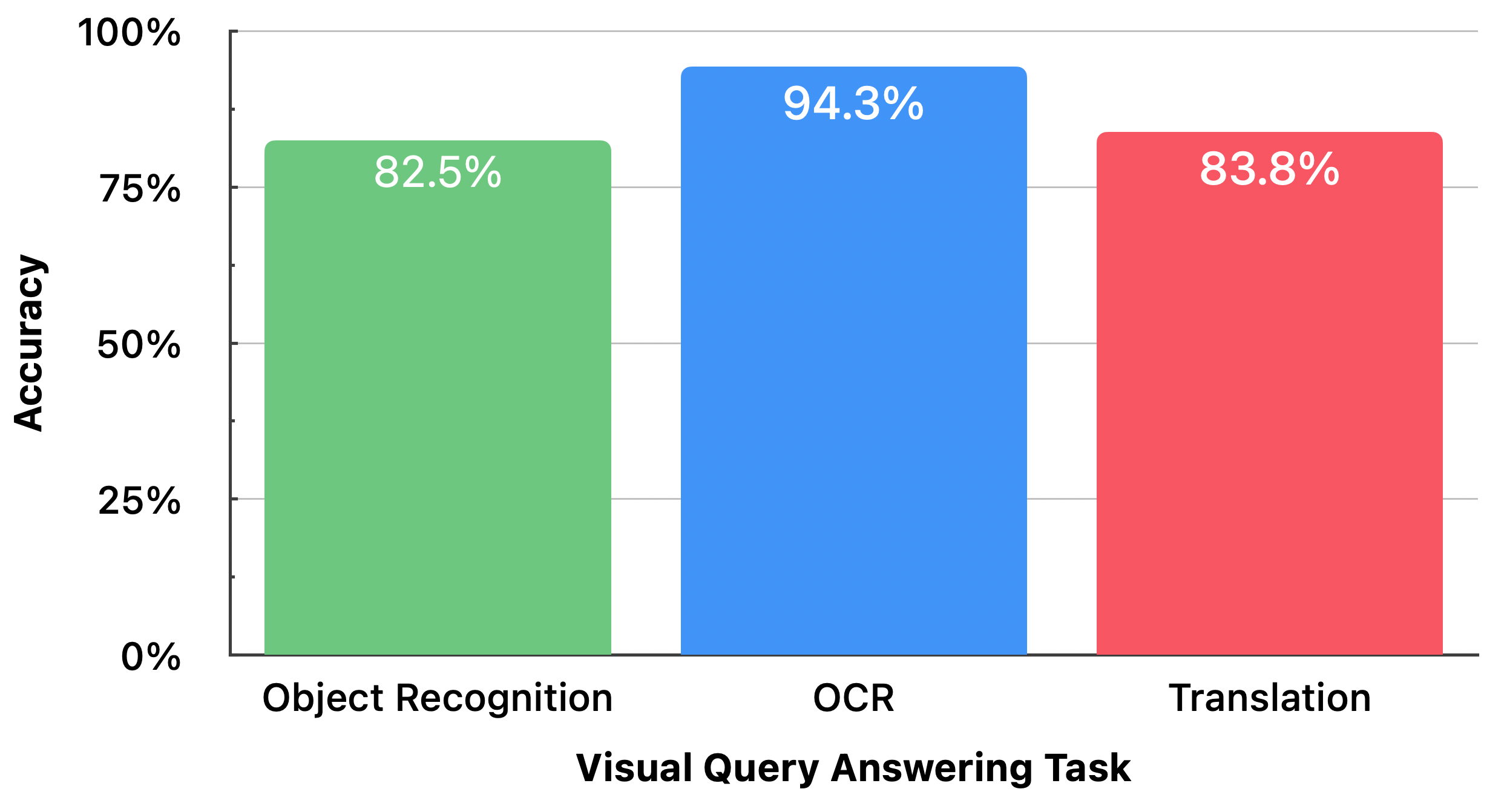}
\vskip -0.2in
\caption{{\bf In-person user study results.} \textmd {VueBuds performance across the three visual question answering (VQA) tasks.}}
\Description{Bar chart showing the in-person user study accuracies across three visual question answering tasks. Object recognition achieved 82.5\% accuracy, optical character recognition (OCR) achieved 94.3\% accuracy, and translation tasks achieved 83.8\% accuracy.}
\vskip -0.25in
\label{fig:in-person-mos}
\end{figure}

{\textbf{Results.}} Across 130 total trials, we achieved accuracies of 82.5\% for Object Recognition, 94.3\% for OCR, 83.8\% for Translation, and an overall accuracy of 86.9\% (Fig.~\ref{fig:in-person-mos}). Notably, OCR outperformed object recognition, contrary to typical task complexity expectations. We attribute the lower object recognition performance to grayscale imaging limitations and increased glare sensitivity with metallic kitchen utensils. For translation tasks, Korean font stylistic variations primarily impacted OCR accuracy rather than translation quality. Our OCR performance on book titles and authors closely aligns with Qwen's reported Doc VQA benchmark of 96.4\%, demonstrating that wearable camera-integrated earbuds can achieve similar performance in real-world usage scenarios. Finally, we highlight representative trials in Fig.~\ref{fig:user-study-highlights}. These examples demonstrate robust performance despite challenging conditions, including partial facial occlusion (Rows 4 and 5), objects split between both images (Rows 1 and 3), and minor hair interference (Rows 5 and 6), validating the system's real-world applicability.

\vskip 0.05in\noindent{\bf Post-Study Feedback.} Participant responses revealed several key themes regarding usability, comfort, and adoption considerations.

\noindent\textit{Physical comfort and device integration:} Most participants reported that VueBuds felt similar to regular earbuds, with one noting they "felt the same way" and another stating "in terms of how it feels on the ears, it feels like wearing normal earbuds." Several participants mentioned the devices were "maybe a little heavier" but emphasized this was "not too noticeable" and wouldn't affect daily use. We note that the weight of the Sony Earbuds at baseline (8.5 grams) is approximately twice that of AirPods (4.3 grams) \cite{sonyearbudspecs, appleairpods4}.

\noindent\textit{Initial thoughts and camera concerns:} Participants had predominantly positive initial reactions, with several expressing enthusiasm: "a very fascinating idea because its very similar to the meta glasses but probably more convenient," "I didn't really have any concerns! Thought it was a pretty cool idea," and "I thought they were cool!" One participant appreciated a unique privacy advantage, noting "I think I could put my hair down and I can stop it from recording any visual data. Seems like that can't be done with glasses." Concerns centered on privacy regarding cloud data streaming, anticipated bulkiness, and camera positioning accuracy.

\noindent\textit{Wearer comfort and privacy:} Participants expressed generally high comfort levels with wearing VueBuds in social contexts, with two responding "very comfortable" and one rating comfort as 7/10. Visual discreteness emerged as important, with one participant appreciating that camera-integrated earbuds "don't have much difference than traditional earbuds looks wise." However, privacy concerns affected comfort levels, with one participant noting "an increased concern in privacy and when the camera is active or what the camera sees." Comfort was often conditional on utility, depending on whether the device "helps me ask questions about the visual scene rather than just taking pictures."

\noindent\textit{Comparison to smartphone alternatives:} Participants overwhelmingly favored VueBuds over phone workflows, describing them as providing "a lot more seamless and integrated experience, less steps." Participants highlighted advantages during physical activities: "when I'm running or biking, it's a lot easier to just say a wake word and ask a question than to stop, take out my phone, take a photo, etc." However, one participant made nuanced distinctions, preferring earbuds for "questions that able to be answered within [a few] seconds" but favoring ChatGPT for "problems that require back and forth conversation or additional details."

\noindent\textit{Practical use cases:} Translation emerged as the most prominent application, particularly for Asian market shopping where participants found it "hard for me to read what it says and the pictures are not sufficient." Travel applications were frequently cited for "traveling in a foreign country with language barriers," along with accessibility support, plant identification, hands-free multitasking during "typing, piano, cooking," and academic support. However, social considerations emerged, with one participant noting voice commands can be "awkward to use in public," suggesting "gestures or nonverbal cues" as potential alternatives.

\subsection {System Evaluations}

\begin{table}[t!]
    \caption{\textmd{VueBuds power consumption results.}}
     \Description{Table showing the VueBuds hardware power consumption by component in each operating state: OFF, IDLE, and ACTIVE. The estimated total power consumption for OFF, IDLE, and ACTIVE are {{34.03 microwatts}}, 3.76 milliwatts, and 21.2 milliwatts respectively. The measured total power consumption for SLEEP, IDLE, and ACTIVE are 86.5 microwatts, 6.63 milliwatts, and 26.1 milliwatts, respectively. }

     \vskip -0.15in     
{
     \begin{tabular}{c c c}
      \hline
      \textbf{Component} & \textbf{IDLE} & \textbf{ACTIVE}\\
      \hline
      SoC (ISP1807) & 3.53 mW & 19.2 mW \\
      PMIC (MAX77650) & 0.148 mW  & 0.148 mW \\
      Camera (HM01B0) & 0.2 mW & 1.1 mW\\
      \hline
      \textbf{Estimated Total} & 3.878 mW & 20.5 mW\\
      \hline
      \textbf{Measured Total} & 3.8 mW & 24.9 mW \\
      \hline
    \end{tabular}
}
    
    \label{tab:power_consumption}

  \hfill
  \vskip -0.15in

  \end{table}
  

\subsubsection {Power analysis}\label{sec:poweranalysis}
To evaluate the viability of VueBuds from a power perspective, we conducted comprehensive profiling across varying usage patterns and analyzed battery life impact.

\begin{figure*}[t!]
\centering
\includegraphics[width=0.7\linewidth]{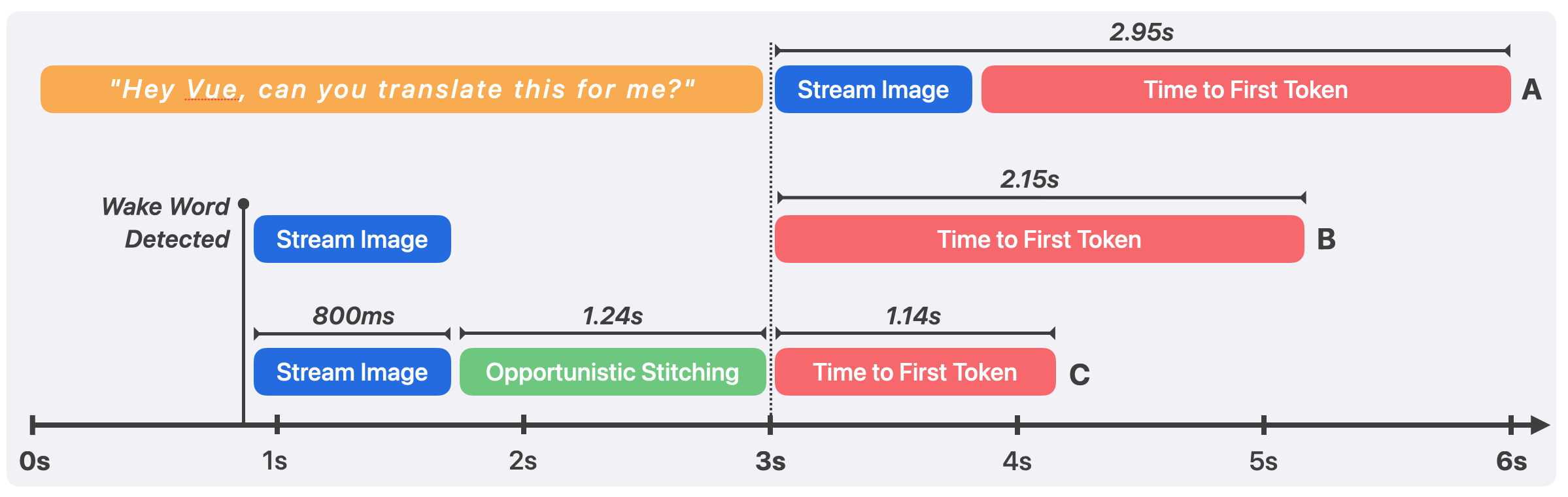}
\vskip -0.1in
\caption{{\textbf{Latency breakdown across three system configurations.} \textmd{(A) Current prototype: 2.95s total latency due to waiting for complete audio processing before camera activation. (B) With on-device wake word detection: 2.15s latency by streaming images during query completion. (C) With opportunistic stitching: 1.14s latency through parallel image stitching while the user speaks, reducing VLM input tokens by 47\%. The vertical dotted line indicates typical query completion time.}}}

\Description{This image shows the latency breakdown for VueBuds across three system configurations when responding to the query “Hey Vue, can you translate this for me?”. Configuration A streams the image and begins inference, resulting in 2.95 seconds to first token. Configuration B processes dual images directly, achieving 2.15 seconds to first token. Configuration C applies opportunistic stitching before inference, reducing the total time to 1.14 seconds.}
\vskip -0.15in
\label{fig:latencyscenarios}
\end{figure*}

\textbf{Power Profiling Methodology.} We connected a power supply with $\mu$A resolution to the battery terminals of the VueBuds camera module, and measured the current draw at 3.8V in both the ACTIVE and IDLE state for at least 20 seconds. For each operating mode, we averaged the current draw and derived the power consumption numbers  in Table \ref{tab:power_consumption}. Based on these measurements, we calculated VueBuds' power consumption across different query frequencies ranging from standby operation (0 queries/hour) to intensive use (60 queries/hour). Each query involves transitioning from IDLE (3.8mW) to ACTIVE (24.9mW), capturing and streaming visual data for three seconds, and then returning to IDLE. We calculated a weighted average power consumption across each minute, accounting for the proportion of time spent in each state. From this analysis, we report that in IDLE mode, the VueBud camera module adds  3.8mW to the baseline power consumption, while intensive usage of 60 queries per hour adds 4.9mW to the baseline.

\textbf{Battery Life Impact on Commercial Earbuds.} We model the battery life impact when integrated with two commercial wireless earbuds: Sony WF-1000XM3 (65mAh battery capacity) ~\cite{SonyBattery} and Apple AirPods Pro 2 (49.7mAh capacity) ~\cite {AirPodsBattCap, AirPodsBattCap2}. Both the devices advertise a 6-hour battery life during streaming with ANC enabled. With the advertised battery life and battery capacity, we calculate the baseline power consumption of both earbuds, which comes out to 40.08mW and 30.65mW, respectively. We then combine this baseline power consumption with our system, and the battery life impact was 11-14\%, even under high usage (60 queries per hour). Table~\ref{tab:batt_life} shows the battery life numbers for VueBuds using the Sony WF-1000XM3. For AirPods Pro 2, we estimate 5.23 and 5.18 hours of battery life at 5 and 60 queries per hour, respectively.


\begin{table}[h!]
\vskip -0.1in
    \caption{\textmd{VueBuds battery life across queries per hour.}}
    
    \Description{Table showing the effect of different numbers of queries per hour on battery life of VueBuds when integrated with the Sony WF-1000XM3 earbuds system. As the number of queries per hour increases, the battery life decreases accordingly. While the VueBuds system is in the user's ear, at 0 queries/hr the battery life is 5.48 hours. At 60 queries per hour, the battery life is reduced to 5.35 hours.}
    
    \vskip -0.15in
 {   
    \begin{tabular}{c c}
      \hline
      \textbf{Queries/hr} & \textbf{VueBuds on Sony Earbuds}  \\
      \hline
      0 & 5.48 hrs  \\
      5 & 5.47 hrs  \\
      10 & 5.46 hrs  \\
      20 & 5.44 hrs  \\
      60 & 5.35 hrs  \\
      \hline
    \end{tabular}
  }  
    \label{tab:batt_life}

  \vskip -0.1in
\end{table}

\textbf{Always-on Camera Analysis.} While wake-word activation suffices for visual question answering, we also evaluated continuous camera operation for potential applications like ambient visual intelligence. With VueBuds constantly in ACTIVE state, battery life decreases to 3.5 hours (Sony WF-1000XM3) and 3.18 hours (AirPods Pro 2), reductions of 42\% and 47\% respectively. Although significant, we note that emerging earbuds like the AirPods Pro 3 ~\cite{AirPodsPro3} advertise 8-hour battery life, suggesting that advances in silicon may  enable always-on visual capture while maintaining acceptable battery life. {Regarding thermal safety, continuous camera operation increases baseline power consumption from 40mW to 65mW, which remains well below thresholds associated with thermal discomfort for on-body wearables~\cite{COMSOLBlog2016}.}  We note that Ray-Ban Meta does not currently support an always-on camera~\cite{safelywearmetarayban}


\subsubsection{Latency Analysis}\label{sec:latencyresults}
To evaluate VueBuds' real-time performance, we define end-to-end latency as the time from query completion ($t_0$) to first token generation by the vision language model ($t_1$). We start off by characterizing our dual image acquisition latency, and then compare the time to first token using dual images versus stitched images. Finally, we analyze three system configurations: our prototype implementation, a system with on-device wake word detection, and an optimized system that additionally performs opportunistic image stitching.

\textbf{Image Acquisition Latency.} We characterize our dual image acquisition latency by measuring the time it takes for the host to trigger the VueBuds cameras over BLE, and receive images from each earbud. Our pipeline sends a BLE write command to both VueBuds to transition them from IDLE to ACTIVE, captures a single camera frame, and transmits the data back to the host. Across 66 trials, we report dual-image acquisition latency as ($\bar{x}=800.1ms$, $\sigma=0.06ms$), which closely aligns with our expectations from \xref{subsec:wireless_latency}.

\begin{figure*}[t!]
\vskip -0.1in
\centering
\includegraphics[width=0.95\linewidth]{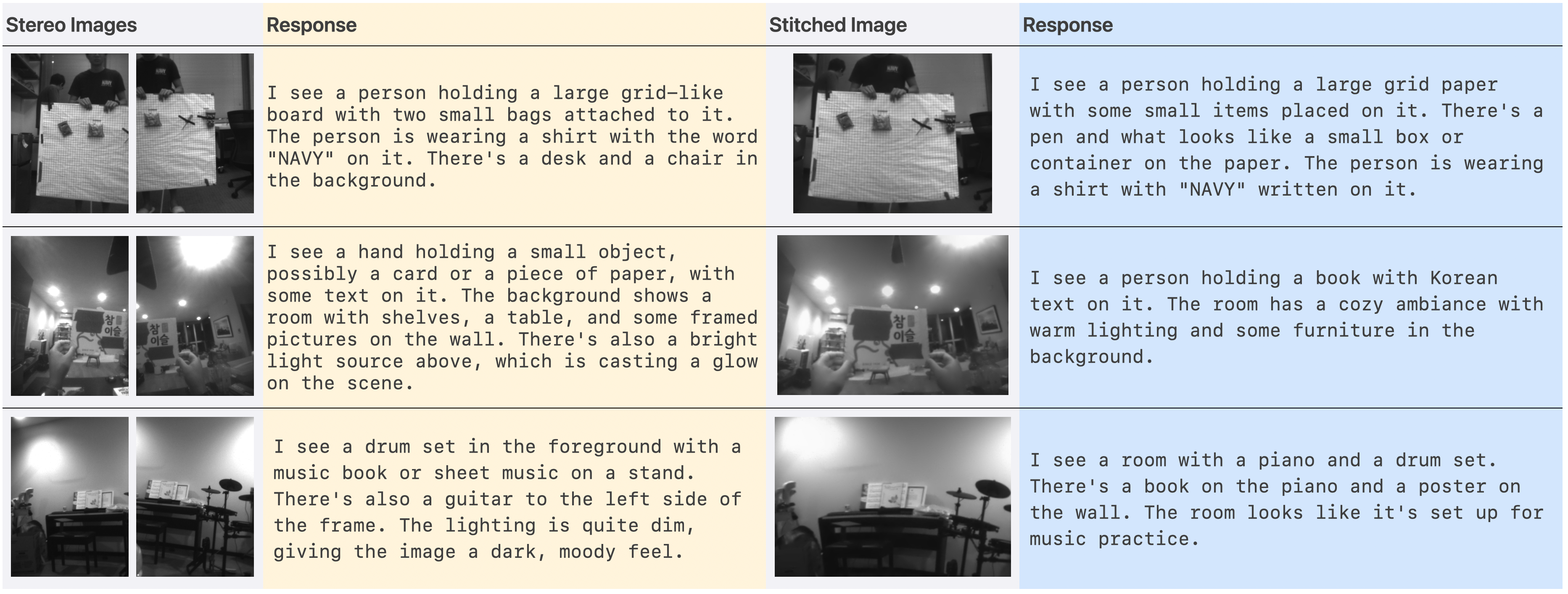}
\vskip -0.15in
\caption{{\bf Comparison of VLM responses to dual images (left) versus stitched panoramic images (right).} \textmd {Reducing redundant information reduces the total input tokens, reducing inference latency by 47\%. } }
\Description{The table presents three rows of visual examples to compare the raw stereo images captured by a device with the final stitched image and the corresponding AI-generated description for each. The table has four columns: 1)  Stereo Images: This column displays two separate images (left and right views) taken from a stereo camera setup. 2)  Response: This column provides the AI-generated description based on the stereo images. 3)  Stitched Image: This column shows the single, combined panoramic image created by stitching the two stereo images together. 4)  Response: This column provides the AI-generated description based on the stitched image.

Example 1: Grid Board

Response (Stereo): "I see a person holding a large grid-like board with two small bags attached to it. The person is wearing a shirt with the word 'NAVY' on it. There's a desk and a chair in the background."

Response (Stitched): "I see a person holding a large grid paper with some small items placed on it. There's a pen and what looks like a small box or container on the paper. The person is wearing a shirt with 'NAVY' written on it."

Example 2: Korean Book

Response (Stereo): "I see a hand holding a small object, possibly a card or a piece of paper, with some text on it. The background shows a room with shelves, a table, and some framed pictures on the wall. There's also a bright light source above, which is casting a glow on the scene."

Response (Stitched): "I see a person holding a book with Korean text on it. The room has a cozy ambiance with warm lighting and some furniture in the background."

Example 3: Music Room

Response (Stereo): "I see a drum set in the foreground with a music book or sheet music on a stand. There's also a guitar to the left side of the frame. The lighting is quite dim, giving the image a dark, moody feel."

Response (Stitched): "I see a room with a piano and a drum set. There's a book on the piano and a poster on the wall. The room looks like it's set up for music practice.".}
\vskip -0.15in
\label{fig:stitching_examples}
\end{figure*}

\textbf{Time to First Token Latency - Dual and Stitched.} To calculate time-to-first-token for dual 324x239 images, we benchmarked multiple trials from our user study data offline. After preloading the model into memory with a dummy query, we measured VLM inference time using paired images with a fixed text query. Direct processing of dual images yielded an average time-to-first-token of 2.15 seconds ($\sigma=0.002$). We then evaluated stitching runtime by inputting these images through our stitching algorithm, yielding an average of 0.123 seconds ($\sigma=0.01$). Finally, we measured inference across the set of successfully stitched images, which achieved an average time-to-first-token of 1.14 seconds ($\sigma=0.23$). This represents a 47\% reduction compared to dual image processing. 

\textbf{End-to-End Latency.} Our current implementation processes audio using  Whisper~\cite{whisper}, which operates on 2.3-second chunks before performing fuzzy matching for wake word detection.  The full pipeline consists of: (1) audio transmission via Bluetooth HFP, (2) wake word detection from buffered audio, (3) BLE command transmission to activate cameras, (4) image capture and streaming, (5) VLM inference, (6) text-to-speech conversion, and (7) audio playback through earbuds. This implementation using Whisper introduces significant latency, as it processes 2.3~s audio chunks and wake word detection occurs after the user has completed their query and not at the onset (Fig.~\ref{fig:latencyscenarios}). 

Existing wake word detection models on embedded devices achieve sub-100ms detection latency from the end of the wake word~\cite{Jose2020}. With proper integration, wake word detection would be integrated with VueBuds hardware, enabling immediate camera activation upon hearing "Hey Vue" or "VueBuds." This would allow images to start streaming while the user completes their query, masking image capture latency. By the time the query ends, the host device has already received and buffered visual data. This optimization would reduce the effective end-to-end latency to the VLM's time to first token (2.15s on M4 Mac Mini Pro) plus audio synthesis overhead.

We also explore optimizing latency through lightweight ORB-based stitching during query completion. While the user speaks, the host device stitches left and right images to reduce redundant visual information. When successful, this reduces VLM input tokens, yielding a 46\% improvement in time-to-first-token compared to raw L/R image pairs. However, due to parallax effects from the spatial separation of ear-level cameras, stitching confidence can vary based on scene geometry and how VueBuds are worn by a user. We introduce this technique as Opportunistic Stitching, achieving a time to first token latency of 1.14 seconds.

Fig.~\ref{fig:latencyscenarios} illustrates the latency breakdown across all three configurations, while Fig.~\ref{fig:stitching_examples} demonstrates successful panoramic stitching results that enable this optimization. These results show that with proper system integration, one could achieve well under 2-second response times, improving the end-to-end user experience.

\section{Discussion and Future Work}

{As our work presents the first system to integrate cameras and visual intelligence with wireless earbuds, we use this section to outline the limitations of our current prototype and, more importantly, articulate promising research directions. We organize this discussion into three key pillars: (1) viewpoint, deictic grounding, and interaction challenges unique to the ear-level vantage point; (2) social dynamics and privacy implications of the form factor; and (3) hardware and platform considerations for scaling to future low-power systems.}

\subsection{Egocentric capture and interaction}

\vskip 0.05in\noindent{\it {Egocentric viewpoint coverage and alignment.}}
 {In \xref{sec:blind-spot-eval}, we empirically validated that ear-level cameras can effectively capture a user's forward field of view through geometric modeling and real-world measurements, with blind spot lengths remaining well below the Harmon distance at camera orientations up to 10°. Notably, VueBuds achieves a 108° horizontal field of view at 10° offset, comparable to Ray-Ban Meta's 100°. In our in-person user study (\xref{sec:inpersonstudy}), participants were instructed to interact with objects naturally by facing their head toward the target. Across 130 trials, no queries failed due to objects being outside the cameras' field of view.}

\begin{figure*}[t!]
\centering
\includegraphics[width=0.95\linewidth]{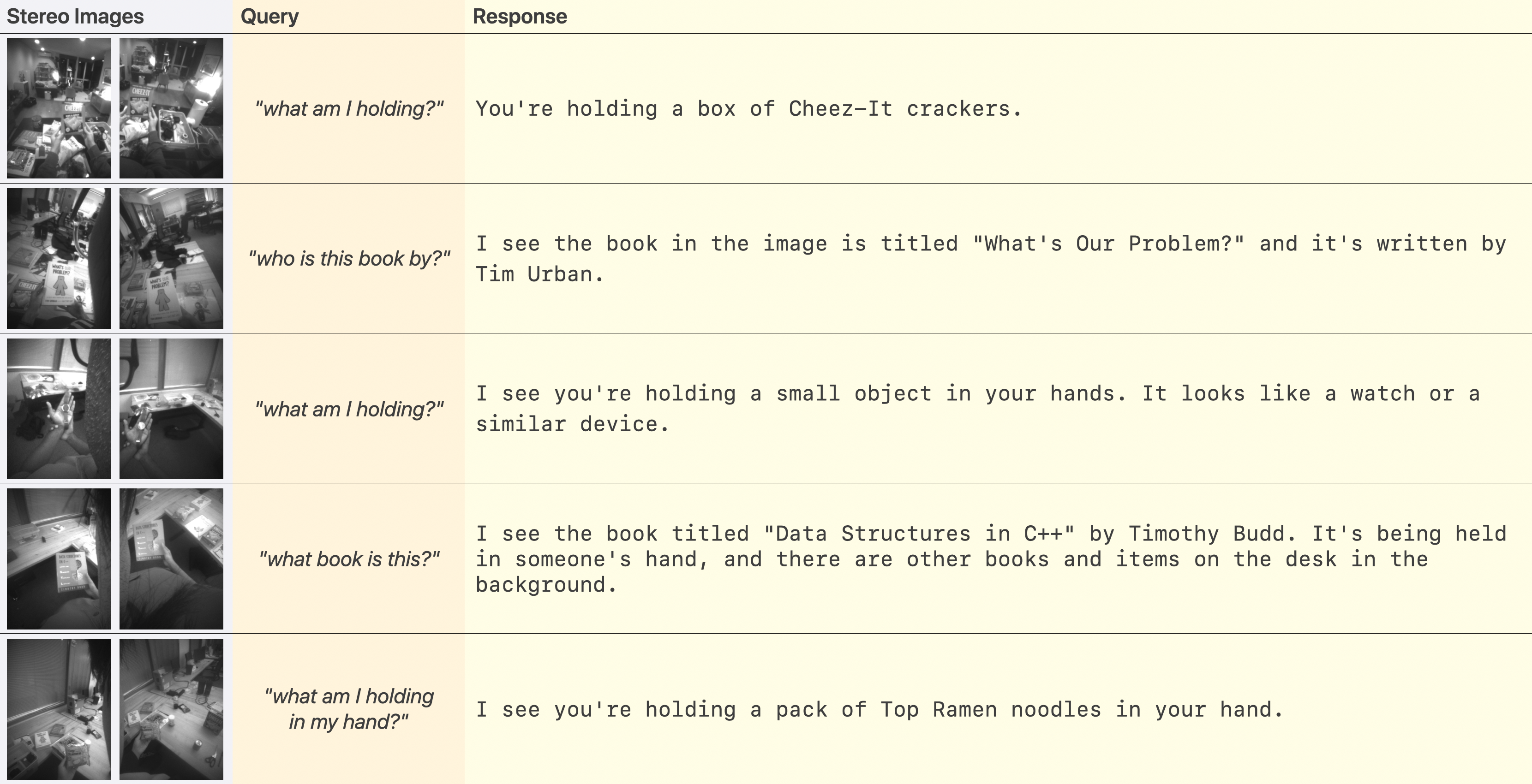}
\vskip -0.1in
\caption{{{\bf Early signs of disambiguation from our in-person study.} \textmd  {Each row shows a cluttered scene with multiple candidate objects. Queries like ``what am I holding?'' or ``what book is this?'' achieve deictic grounding when users grasp the target object.}}}
\Description{Image contains grid of stereo images paired with example queries and responses from VueBuds during in-person user studies. The leftmost column shows grayscale images captured from the earbuds. The middle column contains user queries such as “what am I holding?” or “can you translate this for me?”. The right column shows VueBuds’ responses. Example outputs include identifying a watch, a box of Cheez-It crackers, book identification, and ramen packaging.}
\vskip -0.15in
\label{fig:disambiguation}
\end{figure*}

{Nevertheless, scenarios exist where ear-level cameras may miss a target. While VueBuds provides 87° of vertical coverage, scenarios involving extreme ocular declination without head movement, such as glancing sharply down at a pill bottle while keeping the head level, may fall outside the frame. While the anatomical limit for downward gaze is approximately 47°±8°~\cite{lee2019differences}, ergonomic literature identifies the comfortable resting gaze range as 15-30°~\cite{heuer1989vertical, d2022key}, which sits safely within VueBuds' vertical field of view. This is demonstrated by a sample from our user study in Figure ~\ref{fig:disambiguation} (Row 4) where VueBuds was able to clearly capture and identify an object the user naturally held close to their chest in a cluttered scene. Additionally, earbuds may not be worn consistently across users or sessions, introducing variability in camera alignment. This could be addressed through improved ergonomic design, and future systems could provide auditory feedback if orientation discrepancies are detected between the two earbuds.}

\vskip 0.05in\noindent{\it Deictic grounding and interaction modalities.} {Unlike smart glasses and mixed reality headsets, which can provide both a forward-facing camera and an inward-facing camera for eye-tracking to infer user attention, ear-level cameras represent a different sensing configuration with distinct tradeoffs. Smart glasses can use gaze as an implicit disambiguation signal for referential intent (“look at this”)~\cite{gazepointar,GesPrompt}. In contrast, cameras placed on the ears do not have direct access to the user’s visual fovea or natural gaze cues. }

{Our current VueBuds prototype relies primarily on speech commands for intent expression. Because the system lacks access to gaze, users in our study disambiguated references in cluttered environments by physically picking up and holding objects. This strategy worked well for tabletop interactions but becomes limiting in scenarios where objects are too large, distant, or fixed in place. Compared to systems such as GazePointAR~\cite{gazepointar}, which uses eye tracking for referential grounding, the present VueBuds prototype offers fewer implicit channels for resolving ambiguous queries.}

{Despite these limitations, we see several promising pathways for enhancing disambiguation capabilities in ear-level camera systems. As shown in Figure~\ref{fig:disambiguation}, samples from our in-person study demonstrate that VueBuds can disambiguate objects held in a user's hand even in cluttered visual environments, suggesting that hand-object relationships serve as a strong cue for referential intent. Future systems could extend this capability by using VLMs to infer pointing gestures directly from egocentric imagery, or by incorporating explicit hand tracking to detect pointing direction and identify likely referents. Such pointing gestures have been explored in the context of smart glasses~\cite{GesPrompt}. VueBuds shares a comparable field of view with Ray-Ban Meta, which already supports pointing for referential intent. Incorporating pointing-based grounding, either implicitly via VLM inference or through dedicated hand-tracking, appears well within reach. More broadly, this opens an opportunity for the research community to explore new multimodal disambiguation strategies tailored specifically to ear-level perspectives, rather than adapting techniques built for head-worn devices.}

{Finally, voice-based invocation introduces additional interaction challenges. As noted by a participant in our user study (\xref{sec:inpersonstudy}), voice commands can be ``awkward to use in public,'' with the participant suggesting ``gestures or nonverbal cues'' as potential alternatives. The current single-turn query model further limits recovery from ambiguity, as users cannot engage in multi-turn clarification dialogues to refine their request. Future iterations could explore full-duplex conversational interaction~\cite{veluri2024turnbasedinterfacessynchronousllms} to enable more natural back-and-forth exchanges, as well as alternative input modalities such as in-air gestures recognized by the cameras or existing tap-based mechanisms on the earbud surface.}


\subsection{Social dynamics and privacy}

\begin{figure*}[t!]
\centering
\includegraphics[width=0.95\linewidth]{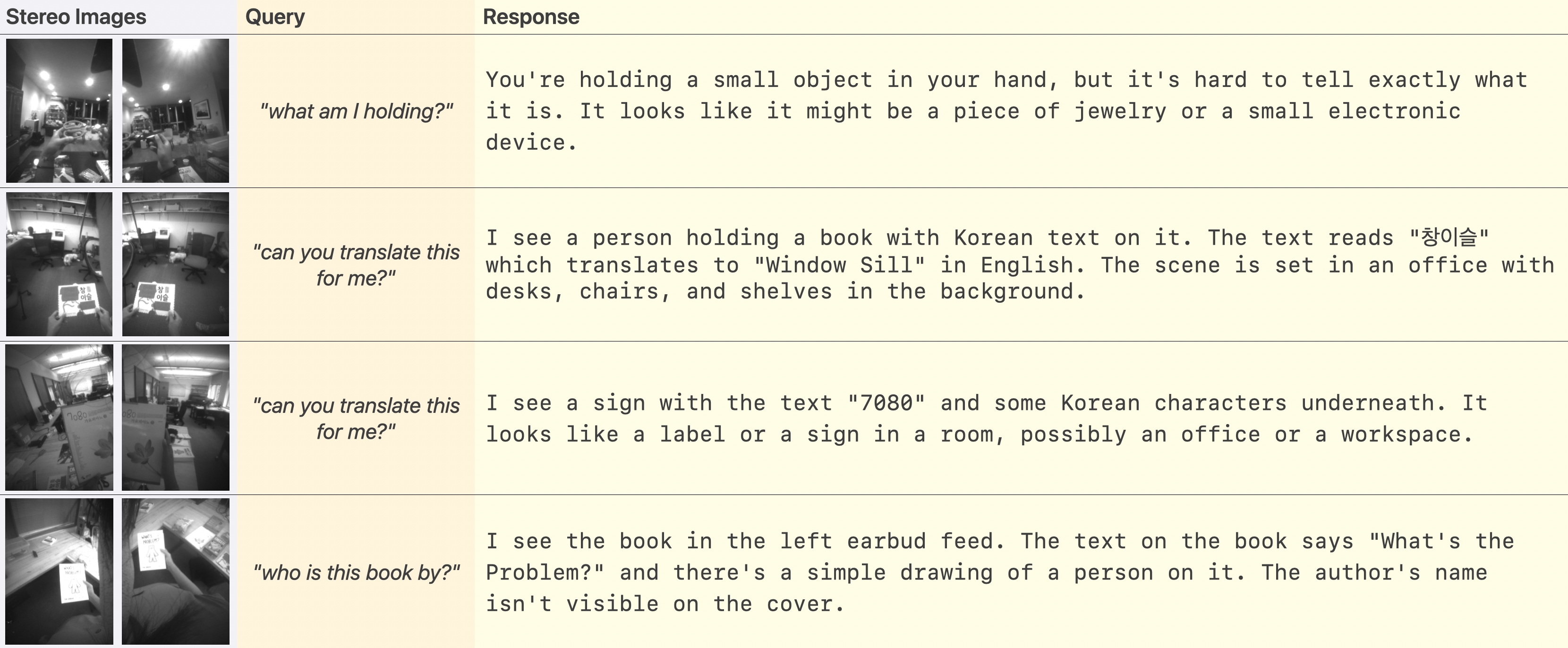}
\vskip -0.1in
\caption{{{\bf Explicit failures from our in-person study.} \textmd {The majority of errors stemmed from hardware imaging limitations. Specifically, low sensor resolution and limited dynamic range compromised OCR performance, particularly on fine print and small text. Additionally, specular glare from adverse lighting conditions disrupted object recognition capabilities.}}}
\Description{Image contains grid of stereo images paired with example queries and responses from VueBuds during in-person user studies. The leftmost column shows grayscale images captured from the earbuds. The middle column contains user queries such as “what am I holding?” or “can you translate this for me?”. The right column shows VueBuds’ responses, specifically failures.}
\vskip -0.15in
\label{fig:failures}
\end{figure*}

\vskip 0.05in\noindent{\it {Privacy signaling and bystander transparency.}}
{Embedding cameras into wearable devices raises privacy concerns for both wearers and bystanders~\cite{bhardwaj2024infocus, iqbal2023adopting}. Extensive prior work on camera-equipped smart glasses highlights concerns and mitigation strategies that are likewise relevant to Vuebuds. Research on smart-glasses cameras found that wearers feel emotionally burdened with preserving bystander privacy, experiencing discomfort and social pressure even when not actively recording~\cite{bhardwaj2024infocus}. When considering ethical implications, users typically compare camera wearables to smartphones and CCTV, and view benign use as acceptable while drawing the line at deliberate concealed recording~\cite{bhardwaj2024infocus}. This presents a challenge for camera-integrated earbuds and glasses which can capture imagery without obvious behavioral cues, compared to smartphones that require a visible holding posture that signals recording intent. Furthermore, prior work found that bystanders do not recognize inconspicuous button taps on smart glasses ~\cite{bhardwaj2024infocus} as recording triggers without verbal audio cues. This complicates the ability of bystanders to provide informed consent ~\cite{iqbal2023adopting}.}

Against this backdrop, VueBuds offers several inherent privacy advantages over existing commercial smart glasses. In contrast to Ray-Ban Meta, all VLM processing for VueBuds occurs on-device, ensuring that captured imagery is never transmitted to the cloud. Additionally, VueBuds captures only low-resolution monochrome images (324$\times$239 pixels) intended for VLM inference rather than photo or video archival, reducing (though not eliminating) the risk of incidental bystander identification. Finally, camera activation explicitly requires a spoken wake-word, providing an audible signal to nearby individuals that visual capture is occurring.

{In an attempt to combat these privacy challenges, smart glasses like Ray-Ban Meta incorporate an outward LED indicator for increased bystander transparency. While an LED is integrated into VueBuds hardware, prior work has shown that they are not fully effective under real-world conditions: they are difficult to see in bright sunlight, unnoticeable at a distance or in crowds, and largely unrecognized by the general public~\cite{bhardwaj2024infocus, iqbal2023adopting}. Audio feedback, such as a shutter sound when capturing, offers the wearer an audible cue that VueBuds could readily support through its integrated speakers. The earbud form factor provides an advantage here, as in-ear audio delivery is more robust against ambient noise compared to the open-ear speakers found in smart glasses. However, while such indicators provide some transparency to wearers, we acknowledge they do not offer complete protection for bystanders, given the current low public awareness of these devices ~\cite{bhardwaj2024infocus}.}

{Several technical mitigations could strengthen privacy in future iterations. Recent work demonstrated that gesture-based image cropping, where users define visual boundaries through hand gestures before VLM processing, reduces explicit privacy leakage (e.g., bystander faces) by 70.6\% while maintaining response quality~\cite{throughlens2025}. Automatic face detection and blurring could further protect bystander identities. Low-tech solutions also merit consideration: physical lens covers, analogous to webcam covers on laptops, provide a mechanism legible to both wearers and observers~\cite{bhardwaj2024infocus}. For VueBuds specifically, as noted by a participant in our user study, wearers can naturally occlude the cameras simply by putting their hair down, providing an intuitive physical control unavailable with glasses-based systems. Prior work has also explored opt-in and opt-out gestures enabling bystanders to signal recording preferences. However, consent-based approaches do not scale to crowded environments and gestures with negative connotations may cause acquiescence effects where bystanders silently accept privacy infringements~\cite{koelle2018gestures, bhardwaj2024infocus}.}

{Finally, even with careful image cropping, VLMs can infer sensitive attributes from subtle background cues. Prior work found that models correctly inferred location from architectural details, signage, and geographical features in 65.6\% of test cases~\cite{throughlens2025}. Addressing such inferential disclosure, particularly when query targets themselves contain identifying features, remains an open challenge for privacy-preserving VLM interaction on wearables.}

\vskip 0.05in\noindent{\it {Form factor and adoption considerations.}} {A central motivation for VueBuds is the observation that wireless earbuds represent a more broadly accessible platform than smart glasses (\xref{sec:platform-accessibility}). For wearers, earbuds benefit from social normalization that smart glasses have struggled to achieve. The negative reception of Google Glass illustrates how prominent camera-equipped eyewear can trigger social friction, referred to as the "Glassholes" effect \cite{due2015social}. In contrast, earbuds are already worn for extended or episodic sessions during commutes, exercise, and work. Participants in our study reported that VueBuds felt similar to regular earbuds, suggesting lower barriers to personal adoption.}

{However, we note that wearer comfort does not resolve bystander concerns. The visual discreteness that makes earbuds socially acceptable to wear may simultaneously make embedded cameras less detectable, and thus more concerning, to bystanders who cannot easily identify when recording is occurring. This tension between wearer accessibility and bystander awareness represents an important design challenge for camera-integrated earbuds, as discussed in the prior section.}

{The earbud form factor does offer one distinct advantage: it complements rather than competes with existing eyewear. Users who require prescription lenses face a particular challenge with camera-equipped glasses: removing the device to protect bystander privacy means sacrificing vision correction. Earbuds, by contrast, are inherently episodic accessories that users routinely remove and stow without functional consequence. For users who depend on prescription eyewear, VueBuds offers visual AI capabilities as a complementary accessory rather than a replacement for something they need.} {The choice between ear-level and eye-level cameras ultimately reflects a trade-off between adoption accessibility and perceptual fidelity. VueBuds prioritizes the former to explore whether useful visual intelligence can be achieved in this more ubiquitous form factor.}

\subsection{Hardware and platform considerations}

\vskip 0.05in\noindent{\it {Resolution, depth, and going beyond greyscale.} }
The  visual question answering tasks evaluated in our user studies did not require very high resolution and 
were color-agnostic, i.e., they were proximal tasks and did not involve identifying specific items by color.  We opted for monochrome imaging to minimize streaming latency, as color imaging would increase bandwidth requirements by 3x (8-bit) or 2.25x (6-bit depth). At our 324x239 resolution, color streaming alone would require 2.4 seconds (8-bit) or 1.8 seconds (6-bit) per frame over our bandwidth-constrained BLE connection, far exceeding acceptable latency for real-time interaction. {However, these constraints introduced concrete failure modes in our in-person study. As shown in Figure~\ref{fig:failures}, the majority of errors stemmed from hardware imaging limitations: low sensor resolution and limited dynamic range compromised OCR performance on fine print and small text, while specular glare from adverse lighting conditions impacted object recognition performance. These failures highlight the tradeoff between streaming latency and resolution in bandwidth-constrained wearable systems.} Future implementations could explore on-device JPEG compression or super-resolution~\cite{neuricam} to reduce bandwidth but support higher resolution, frame rate, and color imaging. {Another noteworthy research direction is integrating wireless time synchronization~\cite{chatterjee2022clearbuds} which could improve stitching success rates. Finally, enabling stereo depth estimation ~\cite{guo2024stereoanythingunifyingstereo, Wang_2023_CVPR} could enhance spatial audio functionality, enabling accurate 3D sound source placement and room acoustic modeling~\cite{10890366}.}

\vskip 0.05in\noindent{ {\it Adaptive camera positioning.} Camera positioning also varies with individual wearing styles, affecting the captured field of view. While our fixed angular orientations (5-10°) generalized well across participants in our user study, individual variations in head/ear anatomy and wearing preferences still introduce some variability. Future systems could incorporate IMU-based pose estimation with micro-servo adjustments~\cite{sciencerobotics_uw} to dynamically adapt camera angles, ensuring optimal perspective regardless of user-specific wearing patterns. 



\vskip 0.05in\noindent{\it Improving panoramic stitching.}  Finally, while our lightweight stitching achieves 46\% improvement in time-to-first-token, traditional panoramic stitching assumes cameras rotate around a fixed point and capture planar scenes, assumptions that are marginally violated by the spatial separation across the head. The resulting parallax effects can exceed homography-based transformation limits, causing our system to fall back to dual independent images. Although parallax-tolerant stitching methods exist~\cite{parallax, parallaxEpipolar, parallexImageStitching}, they require 5-40 seconds of GPU processing. Real-time parallax-tolerant stitching for distributed camera systems remains a research challenge.

\vskip 0.05in\noindent{\it {Computational requirements, deployment trade-offs, and scaling to mobile phones.}} { Our prototype performs VLM inference on desktop-class hardware (Mac Mini M4 Pro). While not available during our development, Apple's FastVLM has since been released \cite{fastvlm2025}. This framework supports the same Qwen model used in VueBuds in its 7B configuration. While the fastest latencies (800–1200ms time-to-first-token on iPhone 16 Pro) currently require the smaller 0.5B configuration, continued advances in mobile hardware and vision encoders suggest that efficient on-device inference with larger models is increasingly feasible.} 

{Additionally, while we used Qwen2.5-VL and achieved promising results, performance could be enhanced through model fine-tuning on domain-specific training data captured from earbud perspectives. Our choice of on-device processing preserves user privacy but incurs latency penalties compared to cloud-based alternatives. Larger cloud models, such as those powering Ray-Ban Meta, would likely provide faster response times and superior performance on complex reasoning tasks, though at the cost of requiring constant connectivity and raising privacy concerns.}

{On the hardware front, while our prototype retrofits existing earbuds, emerging SoCs with integrated Bluetooth Classic and BLE~\cite{qualcomm1,fsc,infeon} 
 could enable custom implementations with tighter integration to better optimize for performance.  Ultra-low-power AI processors like WiseEye2~\cite{wiseeye} demonstrate gesture recognition at 1.2mW, fitting within the power envelope for wireless earbuds. Integrating low-power neural processors could enable on-device preprocessing for in-air gestures for interaction, or human recognition to dynamically control spot-forming radius \cite{soundbubble} or vision-aided speech separation~\cite{AISoundAwareness,semantichearing,lookoncetohear,soundbubble,neuralaids}.}

\section{Conclusion}
This work shows that wireless earbuds can serve as a viable platform for egocentric visual intelligence. By integrating low-power cameras into a familiar ear-worn form factor and pairing them with vision language models, VueBuds enable real-time visual intelligence under strict size, power, and bandwidth constraints. Our results show that binocular ear-level cameras provide comprehensive viewpoint coverage for everyday tasks, and that low-resolution, monochrome imagery is sufficient for scene understanding, text reading, and translation when processed by modern VLMs. Across system benchmarks and user studies, VueBuds achieve response quality comparable to commercial smart glasses while operating within the practical limits of existing earbud hardware. More broadly, VueBuds takes a first step toward bringing visual computing to one of the most widely adopted wearable form factors, positioning earbuds as a viable and broadly accessible platform for visual intelligence.



\balance

\bibliographystyle{ACM-Reference-Format}
\bibliography{references}

\appendix

\providecommand{\qsection}[1]{\subsection*{#1}}
\providecommand{\question}[1]{\vspace{0.5cm}\noindent\textbf{#1}\par}

\providecommand{\pairedresponses}[2]{
    \vspace{0.3cm}
    \noindent
    \begin{minipage}[t]{0.48\linewidth}
        \fbox{\parbox[t]{\dimexpr\linewidth-2\fboxsep-2\fboxrule}{
            \small
            \textbf{Response A:}\\
            \textit{#1}
        }}
    \end{minipage}%
    \hfill%
    \begin{minipage}[t]{0.48\linewidth}
        \fbox{\parbox[t]{\dimexpr\linewidth-2\fboxsep-2\fboxrule}{
            \small
            \textbf{Response B:}\\
            \textit{#2}
        }}
    \end{minipage}
    \vspace{0.2cm}
}


\section{User Study Questionnaire}

\noindent This appendix contains the full questionnaire used in the user study. The study was conducted online using Google Forms.

\qsection{Part 1: Introduction and Platform Accessibility Questions}

\noindent \textit{Introductory Text:} You will be participating in a research study which will ask you about your usage frequency of earbuds and glasses, and then have you rate the overall response quality from a visual AI system on a scale from 1--5.

\question{1. What is your age?}
\begin{itemize}[label=$\square$]
    \item 10--19
    \item 20--29
    \item 30--39
    \item 40--49
    \item 50--59
    \item 60+
\end{itemize}

\question{2. What is your gender?}
\begin{itemize}[label=$\square$]
    \item Male
    \item Female
    \item Non-binary
    \item Prefer not to say
\end{itemize}

\question{3. How often do you wear glasses?}
\begin{itemize}[label=$\square$]
    \item Everyday
    \item 5--6 days a week
    \item 3--4 days a week
    \item 1--2 days a week
    \item I don't wear glasses
\end{itemize}

\question{4. How often do you wear earbuds or headphones?}
\begin{itemize}[label=$\square$]
    \item Everyday
    \item 5--6 days a week
    \item 3--4 days a week
    \item 1--2 days a week
    \item I don't use headphones
\end{itemize}

\qsection{Part 2: Evaluating Visual AI Response Quality}

\noindent \textit{Instructions:} Imagine you are wearing AI-enabled smart glasses. You will be shown a series of images along with a sample question you might ask your glasses. For each image, you will be shown two different AI-generated responses. Please rate each response independently on a scale from 1 (Poor) to 5 (Excellent). Note that the length of the response should NOT be a factor in your rating.

\subsection*{Scenario 1}
\textbf{User Query:} "What do you see" \\
\textit{[Image displayed: Hand holding a book titled "Normal People" by Sally Rooney]}

\pairedresponses%
{The image shows a person holding a book titled "Normal People" by Sally Rooney. The background appears to be a kitchen, with a stove and a pot visible. A checkered cloth hangs on the oven handle.}%
{I see a book titled Normal People by Sally Rooney.}

\subsection*{Scenario 2}
\textbf{User Query:} "What do you see" \\
\textit{[Image displayed: Tray with two coasters, scotch tape, two Nintendo Switch controllers, bag of cookies, and a hair clip]}

\pairedresponses%
{A tray with various items on it. Including nintendo switch controllers, a hair clip, and a bag of snacks.}%
{The image shows a wooden tray with several items on it. There are two Nintendo Switch Joy-Con controllers, a small box labeled "Mikrotik," a small cube-shaped object, and a rectangular object with a pattern that appears to be a comb or a similar item. The objects are arranged in a somewhat scattered manner within the tray.}

\subsection*{Scenario 3}
\textbf{User Query:} "Read any text you can see" \\
\textit{[Image displayed: Book "Normal People" on a sofa]}

\pairedresponses%
{The image shows a book titled "Normal People" by Sally Rooney. The book is partially covered by a textured, fuzzy object, possibly a blanket or scarf. The text is readable and the book appears to be resting on a dark surface.}%
{Normal people, a novel by Salley Rooney}

\subsection*{Scenario 4}
\textbf{User Query:} "Translate this for me" \\
\textit{[Image displayed: Korean Jinro Chamisul bottle with Korean characters on it.]}

\pairedresponses%
{J I N R O C H A M I S U L FRESH SOJUJinro Chamisul}%
{The text on the bottle appears to be in Korean. The readable part says "INRO \dots" which translates to "INRO Chongi Sul" in English. "Chongi Sul" is a type of Korean rice wine.}

\subsection*{Scenario 5}
\textbf{User Query:} "Read any text you can see" \\
\textit{[Image displayed: Book cover "Yellowface" by R.F. Kuang on a sofa]}

\pairedresponses%
{The image shows the cover of a book titled "Yellowface" by R.F. Kuang.}%
{Yellowface R.F. Kuang}

\subsection*{Scenario 6}
\textbf{User Query:} "Translate this for me" \\
\textit{[Image displayed: Korean cream soup food packaging]}

\pairedresponses%
{This appears to be a package of Korean instant noodles.}%
{The text in the image appears to be in Korean. The readable text is "creams" which translates to "Creams" in English.}

\subsection*{Scenario 7}
\textbf{User Query:} "Translate this for me" \\
\textit{[Image displayed: Korean cold noodle dish package on a stovetop]}

\pairedresponses%
{The text on the package appears to be in Korean. The visible text reads "mulnaengmyeon," which translates to "Cold Noodles" in English. This suggests the package contains a cold noodle dish.}%
{The text appears to be a product labeled for a Korean cold noodle dish with ingredients and nutrional info listed.}

\subsection*{Scenario 8}
\textbf{User Query:} "How many screws do you see" \\
\textit{[Image displayed: Five screws on a table]}

\pairedresponses%
{I see several screws on the table.}%
{In the image, there are six screws visible.}

\subsection*{Scenario 9}
\textbf{User Query:} "Read any text you can see" \\
\textit{[Image displayed: Hand holding a Nature Valley chocolate chunk bar]}

\pairedresponses%
{The text on the object in the image reads "Nature Valley."}%
{Nature Valley chewy chocolate chunk.}

\subsection*{Scenario 10}
\textbf{User Query:} "How many calories are in this?" \\
\textit{[Image displayed: Hand holding Nature Valley chocolate chunk bar]}

\pairedresponses%
{The image shows a hand holding a Nature Valley granola bar. The text on the packaging indicates it is a "Nature Valley Granola Bar." To determine the calorie count, you would need to refer to the nutrition label on the packaging, which is not fully visible in this image.}%
{I guess there are around 200 to 250 calories in this nature valley chew chocolate chunk bar.}

\subsection*{Scenario 11}
\textbf{User Query:} "How many calories are in this?" \\
\textit{[Image displayed: Hand holding a can of LaCroix sparkling water]}

\pairedresponses%
{The image shows a can of LaCroix sparkling water. The label on the can indicates it is a "Calorie Free" drink. Therefore, it contains 0 calories.}%
{The can of La Croix sparkling water has zero calories.}

\subsection*{Scenario 12}
\textbf{User Query:} "What do you see" \\
\textit{[Image displayed: Wooden tray with various items, including two nintendo switch controllers, a paintbrush, a pair of scissors, tape, and two coasters]}

\pairedresponses%
{I see a wooden tray with various items on it. I see nintendo switch controlls, a paint brush, and scissors.}%
{The image shows a close-up view of a wooden surface with a rectangular tray containing several items. Inside the tray, there is a small, patterned object that appears to be a coaster or a small mat, a white object with a circular pattern that could be a remote control or a similar device, and a pair of scissors. The scissors are partially open, and the blade is visible.}

\subsection*{Scenario 13}
\textbf{User Query:} "How many calories are in this" \\
\textit{[Image displayed: Can of Spam on a couch]}

\pairedresponses%
{The number of calories in this image is not explicitly stated. However, it appears to be a can of spam which typically contains 180 to 200 calories per serving.}%
{The image shows a can of Spam. To determine the calorie content, you would need to refer to the nutrition label on the can, which is not visible in this image. Spam's calorie count can vary depending on the specific variety and serving size.}

\subsection*{Scenario 14}
\textbf{User Query:} "Read any text you can see" \\
\textit{[Image displayed: Hand holding a can of LaCroix]}

\pairedresponses%
{The text visible in the image reads "La Croix."}%
{Naturally based calorie free.}

\subsection*{Scenario 15}
\textbf{User Query:} "How many brushes do you see" \\
\textit{[Image displayed: Seven Paintbrushes on a coffee table]}

\pairedresponses%
{In the image, there are six paintbrushes visible.}%
{I see seven brushes.}

\subsection*{Scenario 16}
\textit{Note: Throughout this study, you have been rating the response quality between two different AI wearable devices. In these last two examples, we will show you the images from each device as the scene is slightly different between the two.}

\textbf{User Query:} "What kind of plant is this?" \\
\textit{[Image displayed: Fiddle leaf fig]}

\pairedresponses%
{The plant in the image is a fiddle leaf fig.}%
{The image shows a small white dog sniffing a dark green leafy plant in a white pot. The background includes a metallic object, possibly a trash bin or a piece of furniture, and a light-colored floor. The scene is in black and white.}

\subsection*{Scenario 17}
\textbf{User Query:} "What do you see?" \\
\textit{[Image displayed: Cluttered countertop with a can of spam, dog treats, can of LaCroix, Lysol wipes, Lunchables, and a roll of paper towels]}

\pairedresponses%
{The image shows a cluttered countertop with various items. In the foreground, there are a can of beer, a bottle of Lysol disinfectant, a roll of paper towels, and a box of tea or coffee...}%
{I see a kitchen counter with various food items and cleaning supplies including a bag of bone broth bites, a can of spam, a package of lunchables, and a container of lysol wipes.}

\qsection{Part 3: Conclusion}
\vspace{-0.5cm}
\question{Please briefly explain your rating methodology.}
\noindent \textit{[Open-ended text field]}

\question{Please explain why you gave responses a high rating versus a low rating. (Optional)}
\noindent \textit{[Open-ended text field]}

\section{In-Person Follow-Up Survey}

\noindent Thank you for completing our in-person user study! We'd like to follow up on your experience by asking a few questions.

\question{1. How did wearing camera-enabled earbuds feel compared to regular earbuds?}
\noindent \textit{[Open-ended text field]}

\question{2. What were your initial thoughts and/or concerns about cameras in earbuds?}
\noindent \textit{[Open-ended text field]}

\question{3. How comfortable would you be wearing camera-integrated earbuds?}
\noindent \textit{[Open-ended text field]}

\question{4. How does this compare to taking a photo with your phone and asking ChatGPT?}
\noindent \textit{[Open-ended text field]}

\question{5. What situations would you actually use this in?}
\noindent \textit{[Open-ended text field]}

\end{document}